\documentclass[twocolumn]{aastex701}

\usepackage{amsmath}

\newcommand{\Kband}{\textit{K}-band}
\newcommand{\Mband}{\textit{M}-band}

\newcommand{\cisoratio}{\ensuremath{{}^{12}\mathrm{C}/{}^{13}\mathrm{C}}}
\newcommand{\oisoratio}{\ensuremath{{}^{16}\mathrm{O}/{}^{18}\mathrm{O}}}
\newcommand{\ratio}[3]{\ensuremath{{}^{#2}\mathrm{#1}/{}^{#3}\mathrm{#1}}}
\newcommand{\leiden}{Leiden Observatory, Leiden University, P.O. Box 9513, 2300 RA, Leiden, The Netherlands}

\begin{document}

\title{The Isotopic and Elemental Abundances of Planet-Host Star TRAPPIST-1}

\author[0000-0001-9282-9462]{Dar\'io Gonz\'alez Picos}
\affiliation{\leiden}
\email[show]{picos@strw.leidenuniv.nl}

\author[0000-0002-1835-1891]{Ian J.M.\ Crossfield}
\affiliation{Department of Physics and Astronomy, University of Kansas, Lawrence, KS, USA}
\email{ianc@ku.edu}

\author[0000-0002-1221-5346]{David Coria}
\affiliation{Department of Physics and Astronomy, University of New Mexico, Albuquerque, NM, USA}
\email{drcoria@unm.edu}

\author[0000-0003-3667-8633]{Joshua Lothringer}
\affiliation{Space Telescope Science Institute, Baltimore, MD, USA}
\email{jlothringer@stsci.edu}

\author[0000-0002-5258-6846]{Eric Gaidos}
\affiliation{Department of Earth Sciences, University of Hawai'i at M\={a}noa, Honolulu, HI, USA}
\email{gaidos@hawaii.edu}

\author[0000-0001-8782-1992]{Elisabeth A.C.\ Mills}
\affiliation{Department of Physics and Astronomy, University of Kansas, Lawrence, KS, USA}
\email{eacmills@ku.edu}

\author[0000-0003-4760-6168]{Sam de Regt}
\affiliation{\leiden}
\email{regt@strw.leidenuniv.nl}

\author[0000-0002-0845-6171]{Donatella Romano}
\affiliation{INAF-Osservatorio di Astrofisica e Scienza dello Spazio, Bologna, Italy}
\email{donatella.romano@inaf.it}

\author[0000-0003-1624-3667]{Ignas Snellen}
\affiliation{\leiden}
\email{snellen@strw.leidenuniv.nl}

\begin{abstract}

Elemental and isotopic abundances are key tracers of planet formation, stellar evolution, and Galactic chemical evolution.
Very low-mass stars are particularly interesting in this regard, because unlike more massive or evolved stars their photospheric abundances retain the star's natal composition. Cool dwarf spectra have historically been challenging to use for measurements of chemical abundances because of the blending of molecular and atomic features. However, recent advances in molecular line lists, atmospheric models, fitting techniques, and IR spectrographs have enabled the successful measurement of elemental and isotopic ratios in a few dozen very-low-mass stars.
Here, we present near-IR high-resolution spectroscopy of TRAPPIST-1 covering the fundamental and overtone bands of carbon monoxide and its prominent isotopologues. From the joint analysis of the \Kband{} (CFHT/SPIRou) and \Mband{} (Keck/NIRSPEC) spectra, we derive the first stellar C/O ratio and the first carbon- and oxygen-isotope ratios for this star. We obtain a metallicity of $[\mathrm{M/H}]=0.00^{+0.06}_{-0.06}$, a C/O ratio of $0.60^{+0.02}_{-0.02}$, $\cisoratio = 154^{+17}_{-16}$, and $\oisoratio = 490^{+78}_{-64}$. TRAPPIST-1 has generally solar-like elemental abundances, with a $\cisoratio$ that is higher than the solar value and may be modestly elevated relative to some nearby cool dwarfs at similar metallicity. While C/O is most tightly constrained by the \Kband{}, the isotopic detections are driven by the \Mband{} data.

\end{abstract}

\keywords{\uat{Molecular spectroscopy}{2095}, \uat{Infrared spectroscopy}{2285}, \uat{Isotopic abundances}{867}, \uat{Galaxy chemical evolution}{580}, \uat{Late-type dwarf stars}{906}}

\section{Introduction}
\label{sec:intro}

Observations of stellar atmospheres reveal the primitive materials
from which planetary systems form. For example, solar photospheric
abundances of most nonvolatile elements are consistent with those of carbonaceous
meteorites \citep{lodders:2003}. If measured with sufficient
precision, stellar abundances can place strong constraints on the
process of planet formation and even on the composition of a star's
planets \citep{bedell:2018}. Studies of exoplanets, brown dwarfs
and low-mass dwarf stars have advanced considerably with the rapid
developments in infrared instrumentation, from early IR spectroscopic
inferences of stellar properties \citep{mould:1976} to recent,
precise measurements of M dwarf radii, temperatures, metallicities,
and planet properties \citep{mann:2013b,mann:2014,newton:2014,newton:2015,martinez:2017,pavlenko:2002, tsuji:2016,kesseliRadii88Subdwarfs2019,souto:2017,souto:2018,hejazi:2023,hejazi:2024}. Such studies now
use standard atomic and molecular features to extract desired
parameters even from medium-resolution, moderate-S/N data.  High-resolution infrared spectroscopy has more recently made it possible to measure {multiple isotopic abundances} (e.g., \cisoratio{} and
\oisoratio{}).  Though previously measured in many bright giant stars, these isotopic abundances have more recently been measured in an increasing number of cool dwarf stars \citep{crossfield:2019a,xuan:2024a,picos:2025,wang:2026,grasser:2026}, substellar objects \citep[e.g.,][]{zhang:2021a,zhang:2021b,gandhi:2023,picos:2024,regtESOSupJupSurvey2026,gonzalezpicosCarbonIsotopeRatio2026}, and even in hotter main-sequence stars \citep{botelho:2020,coria:2023,coria:2024}.

Stable isotopic ratios serve as tracers of the environment in which the solar system formed, encoding signatures of the composition and chemical processing of the protosolar nebula \citep{geiss_and_gloeckler1998}. The Sun and the solar-system planets largely inherit this composition \citep[e.g.,][]{clayton:2004,lyons:2018}, which can, however, be altered by subsequent processing, e.g.\ atmospheric escape or cometary delivery \citep{genda_and_ikoma_2008,lefour:2026}.
High-resolution spectroscopy of young
stellar objects, giant molecular clouds, and the interstellar medium (ISM) reveals a wide
range of \cisoratio{} and \oisoratio{} isotopic ratios
\citep{goto:2003,sheffer:2007,smith:2015}. Through Galactic chemical-evolution (GCE) mechanisms, heavier (more neutron-rich) isotopes are preferentially produced and these
ratios should decrease as [Fe/H] increases
\citep{kobayashi:2011,prantzos:2018,romanoEvolutionCNOIsotopes2019, romano:2022}. Thus isotopic abundances could also potentially provide an
independent `atomic clock' to better constrain the ages of cool stars,
substellar objects, and free-floating planetary-mass objects \citep{veyette:2018}.

While measurements of isotopes in individual stars are interpreted in the context of those in star-forming regions and Galactic stellar populations, those of planets must be interpreted in terms of fractionation from the primordial planet-forming reservoir, as represented by the star.
The intrinsic scatter in $\ratio{C}{12}{13}$ among
nearby dwarf stars can be measured reasonably well
\citep{crossfield:2019a,botelho:2020,coria:2023,picos:2025}, but \oisoratio{} abundances are only starting to be revealed in significant numbers \citep{crossfield:2019a,coria:2023,picos:2025}.  Perhaps the most prominently varying
isotopic ratio in the solar system is D/H, but, because the timescale
for stellar D fusion is fairly rapid
\citep[$\lesssim$300~Myr;][]{spiegel:2011}, spectroscopy cannot reveal
the initial D/H ratios of old field stars.  However, because the lowest-mass
stars fuse only via the p-p chain,
the C and O isotopic ratios of these stars should reflect their birth
  abundances and can be directly used to interpret isotopic
measurements of the atmospheres of these stars' planets.

\subsection{Low-mass Stars and GCE}
Cool, low-mass stars dominate the stellar population of the Milky Way and possess main-sequence lifetimes that far exceed the age of the Universe \citep{chabrier:2003,bochanski:2010}. Their intrinsic faintness and molecule-rich spectra, however, long made precise abundance measurements difficult and caused M dwarfs to be largely excluded from Galactic chemical-evolution (GCE) studies \citep{woolf:2005}.

This situation has changed over the last two decades. High-resolution optical analyses first measured Fe and Ti abundances in nearby M dwarfs and enabled metallicity calibrations based on molecular indices \citep{woolf:2005,woolfCalibratingDwarfMetallicities2006}. Near-infrared calibrations and large spectroscopic samples then extended stellar-parameter and metallicity measurements to cooler and fainter stars \citep{rojas-ayalaMETALLICITYTEMPERATUREINDICATORS2012,mann:2013b,mann:2014,birkyTemperaturesMetallicitiesDwarfs2020,hejaziChemicalPropertiesLocal2020,hejaziChemicalPropertiesLocal2022}. Detailed abundance analyses of near-infrared spectra moved the field beyond bulk metallicity: \citet{souto:2017,souto:2018} measured individual abundances in planet-hosting M dwarfs, and \citet{souto:2022} showed that abundances of 14 elements in benchmark M dwarfs agree with those of FGK binary companions to within $\lesssim0.08$~dex and reproduce the canonical [X/Fe]--[Fe/H] trends of warmer disk stars. M dwarfs can therefore now be placed in the same GCE framework as FGK stars, rather than treated only as chemically inaccessible low-mass contaminants, although the intrinsic faintness and molecular spectra still present a challenge for homogeneous stellar surveys \citep{magriniArielStellarCharacterisation2022}.

Isotopic abundances provide a complementary diagnostic to elemental composition. Isotope ratios encode the relative contributions of distinct nucleosynthetic channels and stellar populations across cosmic time \citep{zhangStellarPopulationsDominated2018,romano:2022}. The pioneering work of \citealt{tsujiNearinfraredSpectroscopyDwarfs2014} and \citealt{tsuji:2016} demonstrated the feasibility of detecting CO isotopologues in cool dwarfs using high-resolution \Kband{} spectroscopy. Extending this type of analysis to the \Mband{}, \citealt{crossfield:2019a} reported carbon and oxygen isotopic abundances for the fully convective M-dwarf binary GJ~745AB and argued that its unusually large \cisoratio{} and \oisoratio{} might not be explainable by standard local GCE alone. In this context, measuring the elemental and isotopic composition of the low-mass, fully convective TRAPPIST-1 provides a benchmark for comparing an old, fully convective ultracool dwarf against solar values, local thin-disk stars, GCE predictions, and the growing sample of chemically characterized low-mass stars.

\subsection{TRAPPIST-1}

The star TRAPPIST-1 has become one of the most well-known low-mass stars since the discovery of seven Earth-sized transiting exoplanets around it \citep{gillon:2016,gillon:2017,luger:2017}, with several planets in or near the star's habitable zone. This very late-type star (M8V) has $T_\mathrm{eff} = 2569 \pm 28$\,K, roughly solar metallicity, a radius of $0.1192 \pm 0.0013\,R_\odot$, and an age of $7.6\pm2.2$\,Gyr \citep{gillon:2017,burgasserAgeTRAPPIST1System2017,davoudiUpdatedSpectralCharacteristics2024}.
The star is a transitional member of the thin/thick disk. Although its bulk metallicity and basic properties have been reported previously, its stellar C/O ratio and carbon- and oxygen-isotope ratios have not been measured.

Although observations of the TRAPPIST-1 system have not, so far, revealed any unambiguous signs of planetary atmospheres \citep{espinoza:2025,glidden:2025,ducrot:2025,piaulet:2025,greene:2023,ih:2023,zieba:2023,lincowski:2023,connors:2025,gillon:2025}, future observations may yet be able to measure the
isotopic abundances of common C- and O-bearing molecules in the
atmospheres of the TRAPPIST-1 planets \citep{lincowski:2019}.

We present our spectroscopic data and reduction in Sec.~\ref{sec:data}, our retrieval-based approach for measuring abundances in Sec.~\ref{sec:analysis}, the main spectroscopic and abundance results in Sec.~\ref{sec:results}, and their interpretation in Sec.~\ref{sec:discussion}. We conclude in Sec.~\ref{sec:conclusions}.

\section{Observations and Data Reduction}
\label{sec:data}

\begin{deluxetable*}{l l l l l l l l}[bt]
\tabletypesize{\scriptsize}
\tablecaption{Observations. Integration times (\texttt{itime}) are in seconds. Airmass values give the range at the start, middle, and end of each night's observations.\label{tab:obs}}
\tablewidth{0pt}
\tablehead{
\colhead{UT Date} & \colhead{UT Times} & \colhead{Instrument} & \colhead{Band} & \colhead{Airmass} & \colhead{itime} & \colhead{coadds} & \colhead{\# of frames} }
\startdata
2020/08/03 & 10:27--15:23 & Keck/NIRSPEC & M & 1.32--1.10--1.47 & 8.85 & 1 & 358 \\
2020/08/09 & 11:03--15:23 & Keck/NIRSPEC & M & 1.16--1.10--1.63 & 8.85 & 1 & 424 \\
2020/08/10 & 10:58--15:21 & Keck/NIRSPEC & M & 1.16--1.10--1.65 & 11.80 & 1 & 438 \\
2020/08/04 & 12:04--14:41 & CFHT/SPIRou & K & 1.11--1.12--1.30 & 245.16 & 1 & 17 \\
\enddata
\end{deluxetable*}

\begin{figure*}
    \centering
    \includegraphics[width=0.99\textwidth]{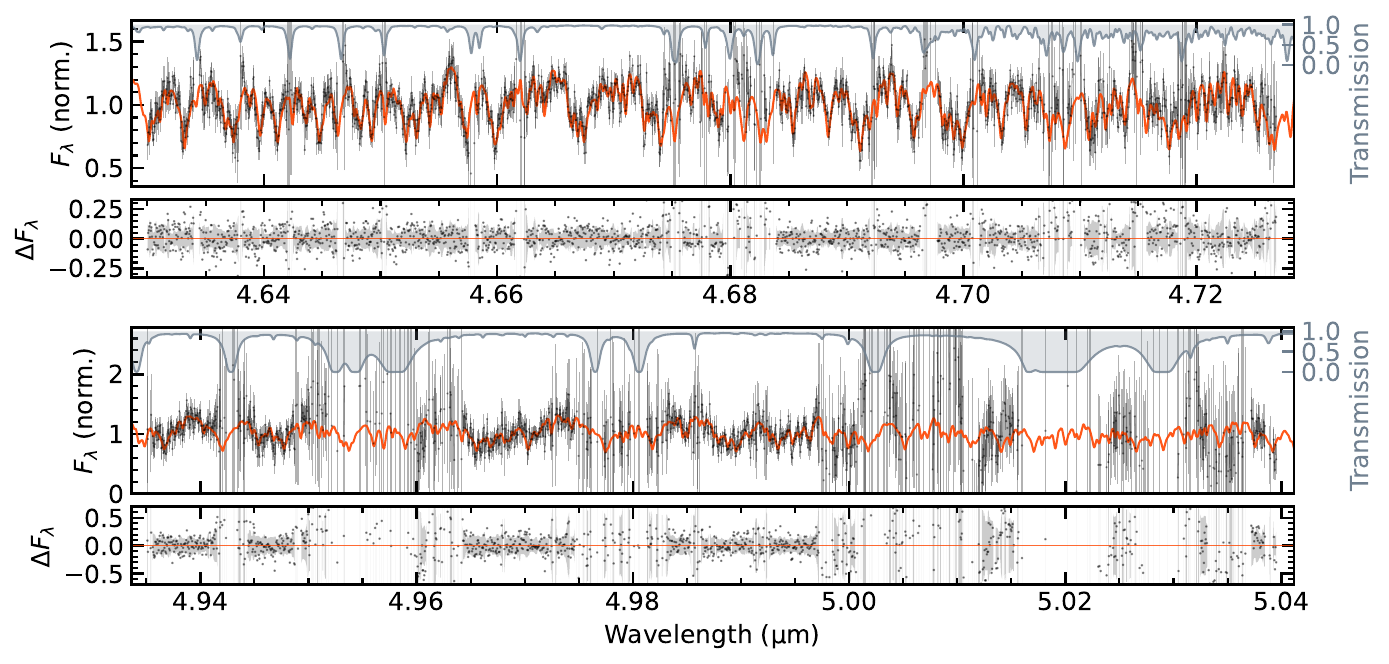}
    \caption{Keck/NIRSPEC \Mband{} spectrum of TRAPPIST-1 (black) with best-fit model (red). \textit{Top:} echelle order 16 ($\lambda\approx4.63$--$4.73$\,\micron). \textit{Bottom:} order 15 ($\lambda\approx4.93$--$5.04$\,\micron). Residuals (data minus model) are shown beneath each panel. Gray shading indicates the relative telluric transmission (dimensionless, 0--1). Masked pixels are omitted from the retrieval.}
    \label{fig:mband}
\end{figure*}

\subsection{\Mband}
We observed TRAPPIST-1 ($m_K=10.3$, $m_{W2}=9.91$) over three half-nights with Keck/NIRSPEC \citep{mclean:1998} as part of program 2020B\_N102 (PI: Crossfield), for a total on-target, open-shutter time of 3.36\,hr.  The instrument was set up in M-wide mode, using echelle position 60.74 and disperser position $\sim$36.7, with the 0.432''$\times$24'' slit. This setting gives useful data over echelle orders 16 and 15, or roughly 4.63--4.73\,\micron\ and 4.93--5.04\,\micron, respectively. We also observed the bright B9V star HR 8865 (95~Aqr; $m_V=5.00$, $m_K=4.96$, $m_{W2}=4.63$) each night for use as a telluric calibrator \citep{vacca:2003}. During all observations we nodded the target star on the slit in an ABBA pattern. Table~\ref{tab:obs} provides further details on our observations, and the raw data are available for download from the Keck Observatory Archive\footnote{\url{https://koa.ipac.caltech.edu/}}. We manually reduced the data and extracted 1D spectra as described in Appendix~\ref{app:nirspec}.  The final spectrum has an S/N in telluric-free regions of roughly 10 and 20~pixel$^{-1}$ in orders 15 and 16, respectively. It is shown in Fig.~\ref{fig:mband}, and is available in machine-readable form in Table~\ref{tab:trappist1_spectrum}.

\begin{figure*}
    \centering
    \includegraphics[width=\textwidth]{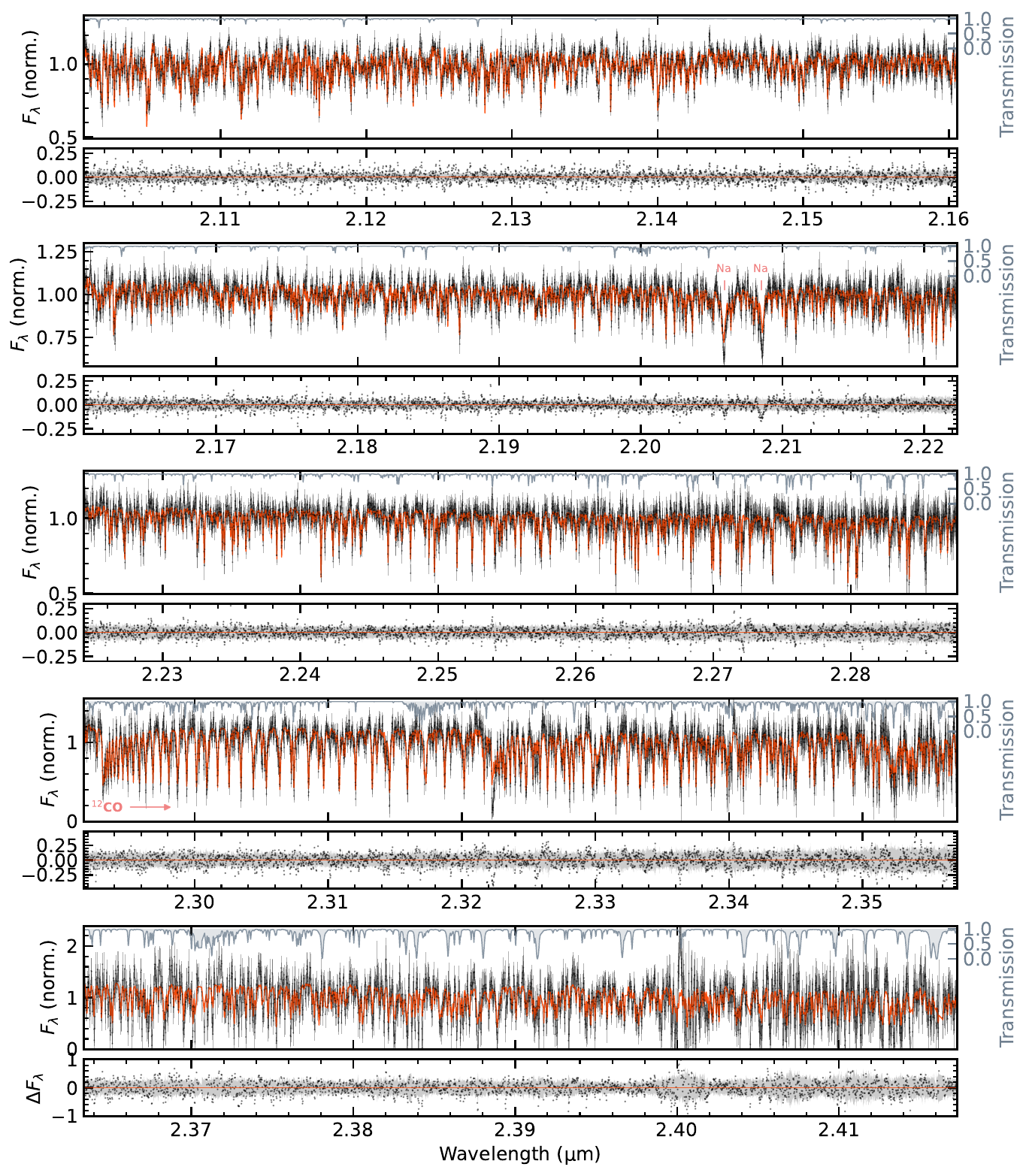}
    \caption{SPIRou \Kband{} spectrum of TRAPPIST-1 (black) with best-fit model (red). H$_2$O dominates the spectrum. The Na doublet appears near 2.206\,\micron\ and strong \({}^{12}\mathrm{CO}\) lines begin near 2.29\,\micron. A nominal telluric transmission spectrum is shown at the top (dimensionless, 0--1). \textit{Bottom:} fit residuals (data minus model).}
    \label{fig:bestfit_spec_spirou}
\end{figure*}

\subsection{\Kband}
We employed archival high-resolution (\(\mathcal{R}\sim70{,}000\)) \Kband{} spectra (2.10--2.42\,\micron) from the SPIRou instrument \citep{donatiSPIRouNIRVelocimetry2020}. Reduced spectra were retrieved from the Canadian Astronomy Data Centre archive and were automatically processed with the SPIRou reduction pipeline (APERO v0.7.28; \citealt{cookAPEROPipelinEReduce2022}), including wavelength calibration and telluric correction.  We combined 17 spectra from the night of 2020 August 4 (PI: Xavier Bonfils, run ID: 20BF18), each with an exposure time of 245.16\,s. The spectra were shifted to correct for barycentric motion, then aligned order-by-order via cross correlation to a reference spectrum at the center of the time series to correct subpixel shifts. We applied sigma-clipping, resampled to a common wavelength grid, and mean-combined the normalized spectra to produce a final spectrum for each order. The uncertainty on the combined spectrum is estimated by calculating the standard error of the mean of the sigma-clipped, resampled spectra ($\sigma_{f_\lambda} = \sigma/\sqrt{17}$) in each wavelength channel. We masked regions affected by strong telluric absorption, identified with a telluric mask by setting lines with depths below 60\% of the continuum to zero. The combined spectrum is shown in Fig.~\ref{fig:bestfit_spec_spirou}.

\section{Analysis}
\label{sec:analysis}

\subsection{Atmospheric Model}
Existing model-atmosphere grids do not span the necessary range of isotopic and elemental abundances, so we computed our own high-resolution emission spectra with \texttt{petitRADTRANS} v3.1 \citep{mollierePetitRADTRANSPythonRadiative2019,blainSpectralModelHighresolutionFramework2024}. We modeled the atmosphere of TRAPPIST-1 using a plane-parallel, hydrostatic framework that computes the emergent spectrum given a temperature structure, volume mixing ratios, surface gravity, and continuum and line opacities.

Line opacities are computed with \texttt{pyROX}\footnote{\url{https://py-rox.readthedocs.io/}} \citep{regtPyROXRapidOpacity2025} using the latest line lists \citep{gordonHITRAN2020MolecularSpectroscopic2022,tennyson2024ReleaseExoMol2024}. Cross sections are computed with pressure broadening for \mbox{H$_2$}-dominated atmospheres following \citet{sharpAtomicMolecularOpacities2007}, and we assume no microturbulence as it is expected to be negligibly small for a 2500 K atmosphere (approximately 0.1 km s$^{-1}$; \citealt{wende3DSimulationsStar2009}).

In this work we use current line lists for the molecular species, namely H$_2^{16}$O \citep{polyanskyExoMolMolecularLine2018}, H$_2^{18}$O \citep{polyanskyExoMolMolecularLine2017}, CO isotopologues \citep{rothmanHITEMPHightemperatureMolecular2010,liROVIBRATIONALLINELISTS2015} and HF, plus atomic data for Na, Ca, and Ti \citep{kuruczModelAtmospheresStars1979}.

Model spectra are generated at resolving power \(\mathcal{R}=500{,}000\) and convolved to the instrument resolution of Keck/NIRSPEC (\(\mathcal{R}=25{,}000\)) and SPIRou (\(\mathcal{R}=70{,}000\); \citealt{donatiSPIRouNIRVelocimetry2020}). We include rotational broadening using an appropriate convolution kernel \citep{grayObservationAnalysisStellar2022} for a given projected rotational velocity ($v\sin i$) fitted as a free parameter.

We parameterize the temperature structure using temperature gradients, with gradients fit to the data \citep{lineUniformAtmosphericRetrieval2015,zhangELementalAbundancesPlanets2023a}. We adopt a similar implementation to \citet{picos:2025}. Here we summarize the key elements. We define the local gradient as \(\nabla \equiv d\ln T/d\ln P\). The temperature at each atmospheric layer \(T_j\) ($n_j=40$) is computed from \(\nabla_j\), obtained by linearly interpolating the gradients \(\nabla_i\) defined at five pressure \emph{nodes} \(P_i\) with \(i=0,1,\ldots,4\):
\begin{equation}
T_j = T_{j-1}\left(\frac{P_j}{P_{j-1}}\right)^{\nabla_j},
\end{equation}
where integration proceeds outward from the anchor \((P_{\rm ref},T_{\rm ref})\). The nodes are fixed at the bottom (\(P_0=100\) bar) and top (\(P_4=10^{-5}\) bar) of the atmosphere. The intermediate nodes are set by the free parameters \(\log P_{\rm ref}\) and \(\Delta \log P = 1.0\), such that \(\log P_{1}=\log P_{\rm ref}-1.0\), \(\log P_{2}=\log P_{\rm ref}\), and \(\log P_{3}=\log P_{\rm ref}+1.0\); the node indices therefore do not correspond to monotonically increasing pressure.

We compute the volume mixing ratios of relevant species assuming chemical equilibrium, using a precomputed \texttt{FastChem} table \citep{kitzmannFastchemCondEquilibrium2023}. We obtain abundances at each pressure--temperature point given the temperature profile, the carbon-to-oxygen ratio (C/O), and the metallicity [M/H]. We define metallicity relative to the solar composition as \([\mathrm{M/H}] = \log\!\left(\frac{Z}{Z_\odot}\right)\), where \(Z=\sum_{i\neq\mathrm{H,He}} N_i/N(\mathrm{H})\) and we adopt solar abundances from \citet{asplundChemicalMakeupSun2021}. In practice, [M/H] is implemented as a uniform scaling factor for all metals except H and He; carbon is then reset at fixed oxygen by \(N(\mathrm{C}) = N(\mathrm{O}) \times (\mathrm{C/O})\), where C/O is fitted as a free parameter. The molecular abundances are then computed in chemical equilibrium using the scaled elemental abundances \citep{kitzmannFastchemCondEquilibrium2023}. To assess detections of molecular and atomic species we compute cross-correlation functions between the observed and model spectra as described in \citet{zhangESOSupJupSurvey2024a}. We also compute the autocorrelation function of the model template to correctly evaluate the cross-correlation S/N.

\subsection{Bayesian retrieval}

We fit the spectra with a Bayesian retrieval that samples nineteen atmospheric and instrumental parameters with \texttt{PyMultiNest} (Table~\ref{tab:trappist1_bestfit}); spline-continuum amplitudes are optimized at each likelihood evaluation rather than sampled (Appendix~\ref{app:spline_continuum}). Priors on the sampled parameters are uniform (Table~\ref{tab:trappist1_bestfit}).

The likelihood uses a diagonal covariance with an uncertainty-inflation parameter \(b\) for each dataset \citep{lineUniformAtmosphericRetrieval2015}. The effective variance for the $i$-th data point is
\begin{equation}
\sigma^2_{\mathrm{eff},i} = \sigma^2_i \times 10^{b},
\end{equation}
where $\sigma_i$ is the formal uncertainty. This multiplicative inflation increases the adopted noise when residuals exceed the formal errors; it does not model correlated noise. We adopt a diagonal \(\boldsymbol{\Sigma}^{-1} = \mathrm{diag}(1/\sigma^2_{\mathrm{eff},i})\) justified by preliminary runs where no significant large-scale correlated residuals are observed.

The chi-squared statistic is computed as:
\begin{equation}
\chi^2 = \mathbf{r}^T\boldsymbol{\Sigma}^{-1}\mathbf{r},
\end{equation}
where $\mathbf{r}$ denotes residuals between the observed and model spectra. The log-likelihood is:
\begin{equation}\label{eq:log_l}
\ln\mathcal{L} = -\frac{1}{2}\left(N\ln(2\pi) + \ln|\boldsymbol{\Sigma}| + \chi^2\right),
\end{equation}
where $N$ is the number of valid data points and $|\boldsymbol{\Sigma}|$ is the determinant of the covariance matrix. To model the continuum of the data we use a flexible spline-based approach as described in \citet{ruffioDetectingExomoonsRadial2023} and implemented in \citet{gonzalezpicosESOSupJupSurvey2025} (see Appendix~\ref{app:spline_continuum}). At each nested-sampling step the spline amplitudes are solved by non-negative least squares and the resulting profile likelihood is passed to MultiNest; we do not analytically marginalize over the continuum coefficients during nested sampling.

We use \texttt{PyMultiNest} \citep{ferozMultiNestEfficientRobust2009, buchnerPyMultiNestPythonInterface2016} to sample the posterior distribution and, when needed for model comparison, estimate evidences. We run retrievals in constant-efficiency mode at 5\% efficiency with 1000 live points and an evidence tolerance of \(\Delta \log Z = 0.5\).

\section{Results}
\label{sec:results}

\subsection{Best-fit spectra and detected species}

\begin{figure*}
    \centering
    \includegraphics[width=\textwidth]{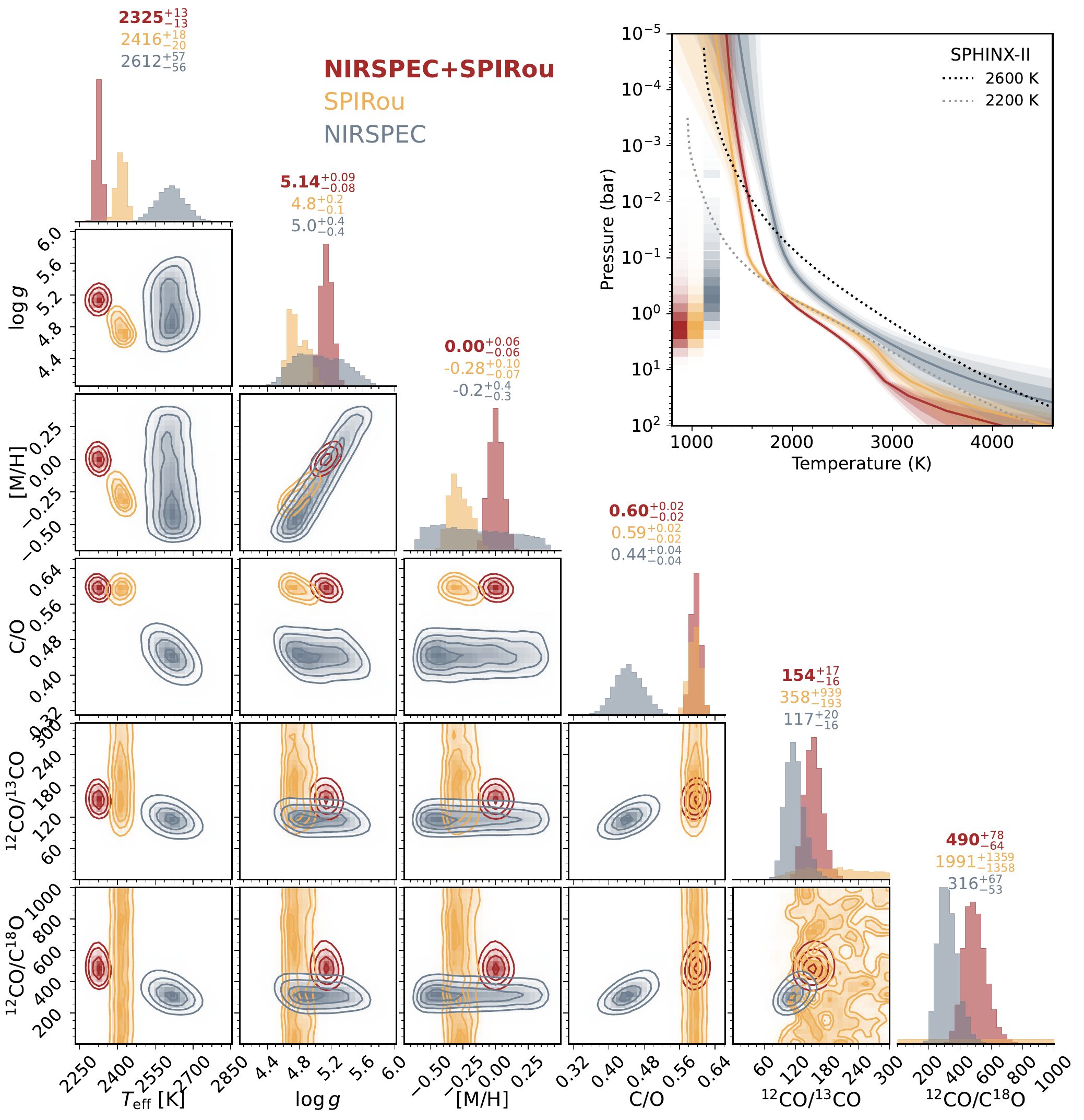}
    \caption{Posterior distributions from joint \Kband+\Mband{} (brown), \Kband-only (orange), and \Mband-only (gray) retrievals. The parameters shown are bolometric effective temperature \(T_{\mathrm{eff}}\) (derived, not sampled), surface gravity \(\log g\), metallicity \([\mathrm{M/H}]\), C/O, and isotopic ratios \cisoratio{} and \oisoratio{} from CO isotopologues. Diagonal panels show marginalized posteriors with medians and 16th--84th percentiles. Off-diagonal panels show joint posteriors with contours enclosing 39\%, 68\%, 86\%, and 95\% of the probability (0.5--2.0\(\sigma\) for Gaussian marginals). \textit{Top:} retrieved pressure--temperature profiles with 68\% and 95\% envelopes; the emission-contribution function is normalized to its peak at each wavelength. SPHINX-II models \citep{iyerSPHINXDwarfSpectral2025} are shown for reference.}
    \label{fig:cornerplot}
\end{figure*}

The joint \Kband{} and \Mband{} retrieval fits the observed spectra well (Figs.~\ref{fig:mband} and~\ref{fig:bestfit_spec_spirou}): most molecular features are matched, with only small residuals in some atomic lines and regions of strong telluric absorption. We find broadly solar composition, confirming earlier near-solar-metallicity measurements, and obtain the first stellar C/O ratio, which agrees with the solar value within our statistical uncertainties. The \Mband{} fit is excellent throughout order 16 and through much of order 15, while strong telluric absorption prevents useful constraints at $\lambda>5\,\micron$. Posterior distributions and pressure--temperature profiles are shown in Fig.~\ref{fig:cornerplot} and Table~\ref{tab:trappist1_bestfit}; the data and model spectra are in Table~\ref{tab:trappist1_spectrum}.

\begin{figure}[t]
    \centering
    \includegraphics[width=\columnwidth]{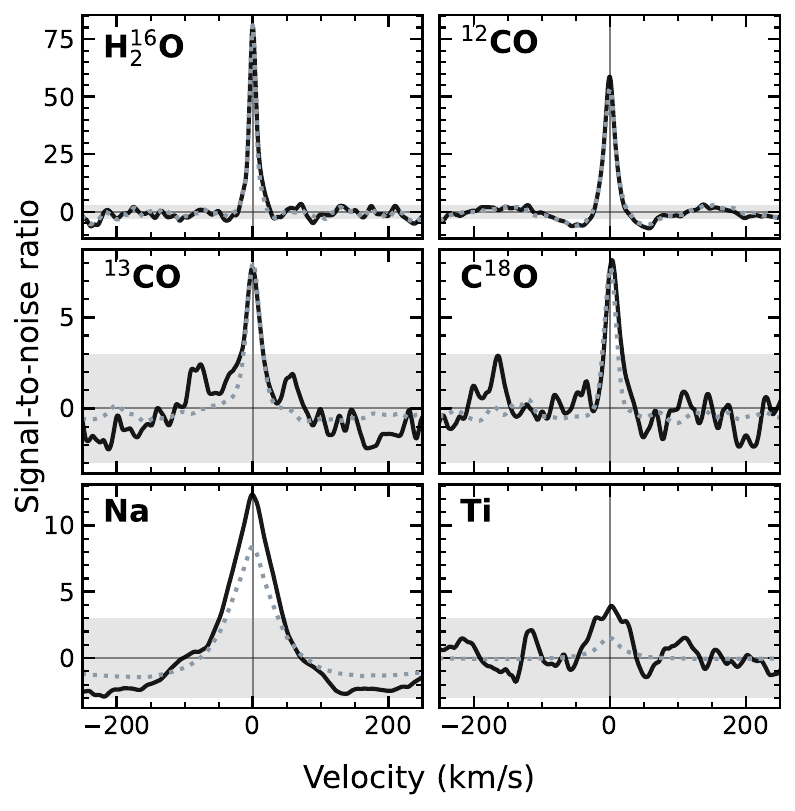}
    \caption{Cross-correlation functions (CCFs) from the joint \Kband+\Mband{} retrieval, evaluated on the residual spectrum after subtracting the best-fit model without the species of interest. Velocity is in km\,s$^{-1}$ relative to the best-fit systemic velocity of each dataset. Species CCFs (black) are normalized by the template autocorrelation (dotted gray); the shaded region marks $|\mathrm{S/N}|=3$. Panels show, from top to bottom: H$_2$O, $^{12}$CO, $^{13}$CO, and C$^{18}$O.}
    \label{fig:cross_correlation}
\end{figure}

\begin{figure*}[htpb]
    \centering
    \includegraphics[width=0.99\textwidth]{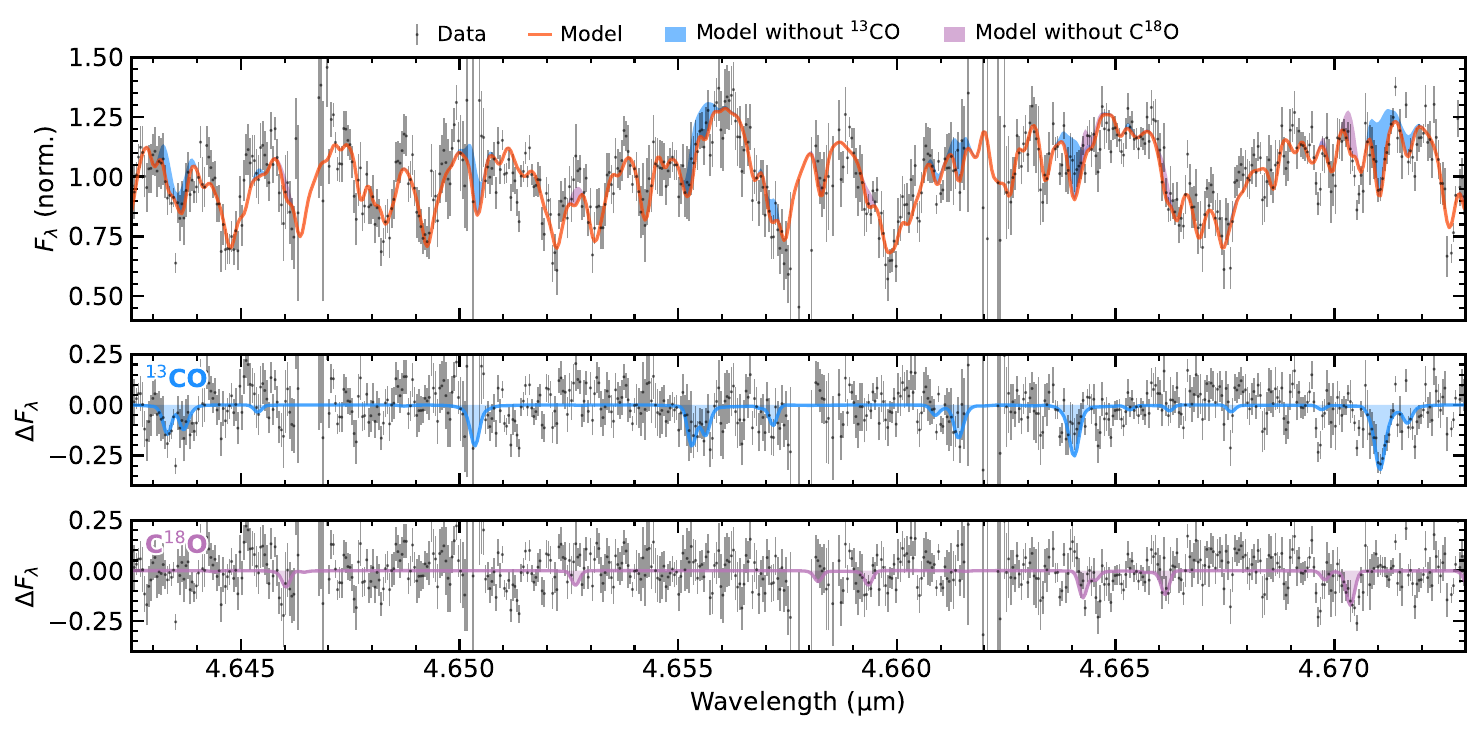}
    \caption{\({}^{13}\mathrm{CO}\) and \(\mathrm{C}{}^{18}\mathrm{O}\) in the \Mband{} (4.642--4.673\,\micron), covering the strongest minor-isotopologue features ($\sim$1/6 of our \Mband{} range). \textit{Top:} observed spectrum (black) with full model (red). Colored curves show the model contributions from \({}^{13}\mathrm{CO}\) (blue) and \(\mathrm{C}{}^{18}\mathrm{O}\) (pink) with all other parameters fixed to the joint best fit. \textit{Middle:} residuals (data minus model) after removing only the \({}^{13}\mathrm{CO}\) opacity. \textit{Bottom:} residuals after removing only the \(\mathrm{C}{}^{18}\mathrm{O}\) opacity. The leave-one-species-out models are not reoptimized; they illustrate how each minor isotopologue contributes to the fit.}
    \label{fig:mband_isotopologues}
\end{figure*}

We detect H$_2$O as the dominant opacity source across the full wavelength range, together with strong \({}^{12}\mathrm{CO}\) lines in the first-overtone band at 2.3\,\micron\ and the fundamental band at 4.6\,\micron. From the \Mband{} spectrum we report detections of the CO isotopologues \({}^{13}\mathrm{CO}\) and \(\mathrm{C}{}^{18}\mathrm{O}\) with $\ln B=136.7$ and $69.1$, respectively. These detections are corroborated by the cross-correlation analysis (Fig.~\ref{fig:cross_correlation}), which is not independent because it uses the same spectra and templates. We quantify detection significance using the log Bayes factor of retrievals excluding the corresponding isotopologue from the model (Fig.~\ref{fig:mband_isotopologues}), following \citet{picos:2025}, and interpret the values with \citet{thorngrenBayesianModelComparison2026}.

No rare CO isotopologues (e.g., $^{13}$C$^{18}$O, $^{12}$C$^{17}$O) are detected in the \Kband{} data. From the \Kband{} spectrum we detect Na, tentatively detect Ti, and report nondetections of OH and HF.

\begin{deluxetable*}{llcc}[h]
    \tabletypesize{\normalsize}
    \tablecaption{Sampled atmospheric and nuisance parameters for the joint Keck/NIRSPEC+SPIRou fit of TRAPPIST-1. \label{tab:trappist1_bestfit}}
    \tablehead{
    \colhead{Description} & \colhead{Parameter} & \colhead{Prior (Uniform)} & \colhead{Posterior}}
    \startdata
    Metallicity & $[\mathrm{M/H}]$ & $[-0.60,\,0.40]$ & $0.00^{+0.06}_{-0.06}$ \\
    Carbon-to-oxygen ratio & C/O & $[0.20,\,0.90]$ & $0.60^{+0.02}_{-0.02}$ \\
    Isotopologue ratio of CO with ${}^{13}\mathrm{C}$ & $\log_{10}({}^{12}\mathrm{C}^{16}\mathrm{O}/{}^{13}\mathrm{C}^{16}\mathrm{O})$ & $[0.0,\,3.6]$ & $2.19^{+0.05}_{-0.05}$ \\
    Isotopologue ratio of CO with ${}^{18}\mathrm{O}$ & $\log_{10}({}^{12}\mathrm{C}^{16}\mathrm{O}/{}^{12}\mathrm{C}^{18}\mathrm{O})$ & $[0.0,\,3.6]$ & $2.69^{+0.07}_{-0.07}$ \\
    Isotopologue ratio of CO with ${}^{17}\mathrm{O}$ & $\log_{10}({}^{12}\mathrm{C}^{16}\mathrm{O}/{}^{12}\mathrm{C}^{17}\mathrm{O})$ & $[0.0,\,3.6]$ & $\geq 3.36$ \\
    Surface gravity & $\log g$ [cgs] & $[4.0,\,6.0]$ & $5.14^{+0.09}_{-0.08}$ \\
    Temperature at $P_{\rm ref}$ & $T_{\rm ref}$ [K] & $[2000,\,3000]$ & $2274^{+29}_{-95}$ \\
    Reference pressure $P_{\rm ref}$ & $\log P_{\rm ref}$ [bar] & $[-1.0,\,1.0]$ & $0.13^{+0.05}_{-0.06}$ \\
    Temperature gradient at $P_{0}$ & $\nabla_{T,0}$ & $[0.08,\,0.42]$ & $0.2^{+0.2}_{-0.2}$ \\
    Temperature gradient at $P_{1}$ & $\nabla_{T,1}$ & $[0.06,\,0.36]$ & $0.062^{+0.003}_{-0.002}$ \\
    Temperature gradient at $P_{2}$ & $\nabla_{T,2}$ & $[0.06,\,0.36]$ & $0.172^{+0.005}_{-0.005}$ \\
    Temperature gradient at $P_{3}$ & $\nabla_{T,3}$ & $[0.02,\,0.22]$ & $0.037^{+0.005}_{-0.006}$ \\
    Temperature gradient at $P_{4}$ & $\nabla_{T,4}$ & $[0.00,\,0.22]$ & $0.01^{+0.02}_{-0.01}$ \\
    Projected rotational velocity & $v \sin i$ [km\,s$^{-1}$] & $[1.0,\,20.0]$ & $2.15^{+0.19}_{-0.19}$ \\
    Radial velocity of SPIRou & $v_{\rm rad}$ (SPIRou) [km\,s$^{-1}$] & $[-90,\,90]$ & $-52.97^{+0.04}_{-0.04}$ \\
    Radial velocity of NIRSPEC order 16 & $v_{\rm rad}$ (order 16) [km\,s$^{-1}$] & $[-90,\,90]$ & $-3.4^{+0.2}_{-0.2}$ \\
    Radial velocity of NIRSPEC order 15 & $v_{\rm rad}$ (order 15) [km\,s$^{-1}$] & $[-90,\,90]$ & $-78.0^{+0.6}_{-0.6}$ \\
    Log$_{10}$ uncertainty inflation & $b$ (SPIRou) & $[-1.0,\,2.0]$ & $0.699^{+0.003}_{-0.003}$ \\
    Log$_{10}$ uncertainty inflation & $b$ (NIRSPEC) & $[-1.0,\,2.0]$ & $0.262^{+0.007}_{-0.007}$ \\
    \hline
    Bolometric effective temperature$^{\mathrm{a}}$ & $T_{\rm eff}$ [K] & \nodata & $2325^{+13}_{-13}$ \\
    Carbon isotope ratio$^{\mathrm{b}}$ & $^{12}\mathrm{C}/{}^{13}\mathrm{C}$ & \nodata & $154^{+17}_{-16}$ \\
    Oxygen isotope ratio$^{\mathrm{b}}$ & $^{16}\mathrm{O}/{}^{18}\mathrm{O}$ & \nodata & $490^{+78}_{-64}$ \\
    \enddata
    \tablecomments{$^{\mathrm{a}}$Derived by bolometric flux integration of the best-fit model over $0.3$--$20$\,\micron\ (see Sec.~\ref{sec:teff}). $^{\mathrm{b}}$Converted from the retrieved log-isotopologue ratios. Posterior intervals are 16th--84th percentiles and represent statistical uncertainties from the flexible one-column retrieval only. The C$^{17}$O posterior piles up at the prior upper bound; we quote a $\sim$99\% lower limit of $\log_{10}({}^{12}\mathrm{C}^{16}\mathrm{O}/{}^{12}\mathrm{C}^{17}\mathrm{O})\geq3.36$ ($\ratio{O}{16}{17}>2300$; $\ratio{O}{18}{17}>4.3$).}
\end{deluxetable*}

\subsection{Elemental and isotopic abundances}

Under our preferred flexible one-column retrieval, the joint fit yields a C/O ratio of \(0.60^{+0.02}_{-0.02}\) and a metallicity of \([\mathrm{M/H}] = 0.00^{+0.06}_{-0.06}\) (Fig.~\ref{fig:cornerplot}). From the individual CO isotopologue abundances we infer
\begin{equation}
\cisoratio = 154^{+17}_{-16} \quad \text{and} \quad \oisoratio = 490^{+78}_{-64}.
\end{equation}

The \Kband-only retrieval anchors C/O and \([\mathrm{M/H}]\), whereas the \Mband-only retrieval drives the isotope ratios (Fig.~\ref{fig:cornerplot}). C/O is stable across the three atmosphere parameterizations tested in Sec.~\ref{sec:two_column} ($0.60$, $0.614$, and $0.58$), but \([\mathrm{M/H}]\) and both isotope ratios vary substantially when the pressure--temperature profile is changed.

\begin{figure*}
    \centering
    \includegraphics[width=\textwidth]{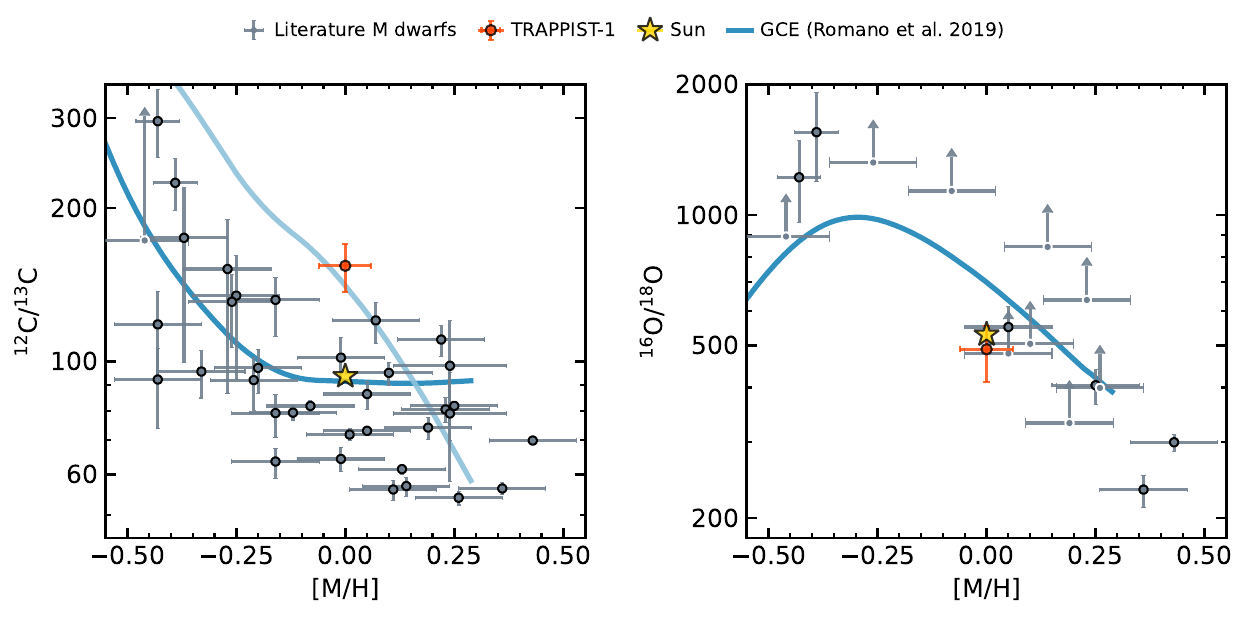}
    \caption{Carbon (\cisoratio) and oxygen (\oisoratio) isotopic ratios of cool dwarfs as a function of metallicity. The horizontal axis shows literature metallicities as quoted by each study and may mix \([\mathrm{M/H}]\) and \([\mathrm{Fe/H}]\) conventions; young, veiled, or substellar systems are excluded. Chemical-evolution tracks \citep{romanoEvolutionCNOIsotopes2019,romano:2022} are shown for different nova-progenitor mass ranges (see \citealt{romano:2022} for details). Literature measurements are from \citet{crossfield:2019a,xuan:2024a,picos:2025,wang:2026} and our own TRAPPIST-1 measurement as reported in this work.}
    \label{fig:isotope_ratios}
\end{figure*}

The oxygen isotopic ratio is consistent with solar system and present-day ISM values within our statistical uncertainties (see Fig.~\ref{fig:isotope_ratios}; \citealt{wilsonIsotopesInterstellarMedium1999}; \citealt{lyons:2018}), while the carbon isotopic ratio exceeds the solar-system value \citep{lyons:2018}. We discuss the interpretation of these abundance patterns in Sec.~\ref{sec:discussion}.

\section{Discussion}
\label{sec:discussion}

\subsection{CO bands as abundance diagnostics}

\begin{figure}
    \centering
    \includegraphics[width=\columnwidth]{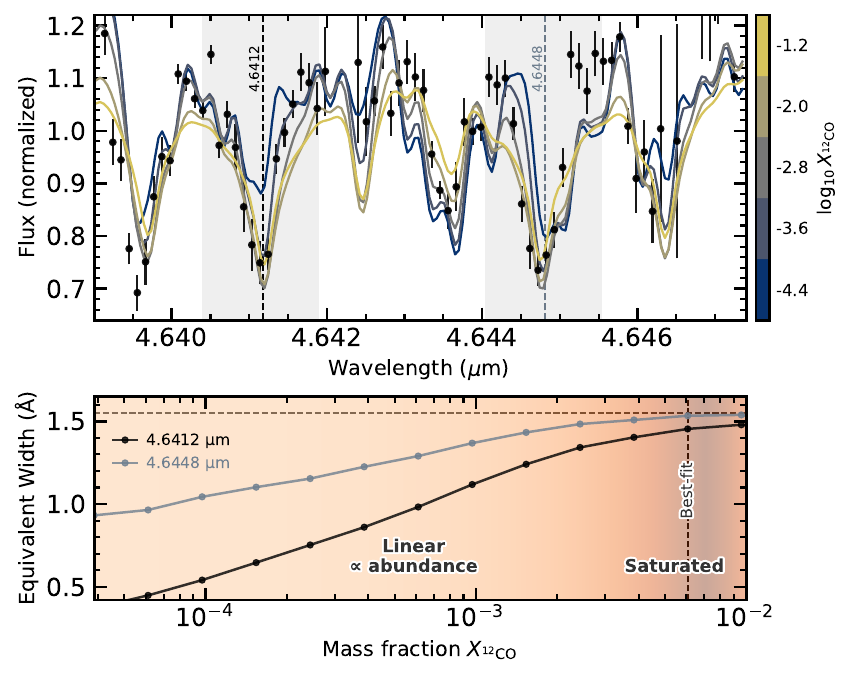}
    \caption{Saturation of \({}^{12}\mathrm{CO}\) in the \Mband{}. \textit{Top:} model spectra for several \({}^{12}\mathrm{CO}\) volume mixing ratios with observed data overplotted (normalized flux versus wavelength in microns). \textit{Bottom:} curve of growth for the line at 4.6412\,\micron\ and the line at 4.6448\,\micron, showing equivalent width (\AA) versus $\log_{10}$ of the \({}^{12}\mathrm{CO}\) abundance. A dashed vertical line marks the retrieved abundance; the observed lines lie in the saturated regime.}
    \label{fig:saturated_12CO_lines_Mband}
\end{figure}

In cool stellar atmospheres, most carbon is stored in molecular CO, which makes CO absorption a precise tracer of carbon abundance \citep{tsujiNearinfraredSpectroscopyDwarfs2014}. At high spectral resolution, individual CO isotopologue lines can also be separated \citep{tsuji:2016,molliereDetectingIsotopologuesExoplanet2019}, enabling the \({}^{13}\mathrm{CO}\) and \(\mathrm{C}{}^{18}\mathrm{O}\) detections reported in Sec.~\ref{sec:results}.
However, the \({}^{12}\mathrm{CO}\) lines in the \Mband{} are saturated (Fig.~\ref{fig:saturated_12CO_lines_Mband}), complicating the inference of isotopic ratios because the \({}^{12}\mathrm{C}\) abundance becomes less sensitive to the \({}^{12}\mathrm{CO}\) line depths. This behavior is captured by the curve of growth (e.g., \citealt{kuruczModelAtmospheresStars1979}, bottom panel of Fig.~\ref{fig:saturated_12CO_lines_Mband}). In the weak-line (optically thin) regime ($\tau_\nu \lesssim 1$, where $\tau_\nu=\int \kappa_\nu \rho\, ds$ is the line-center optical depth), line depths and equivalent widths scale approximately linearly with abundance. As abundance increases, the core of the line saturates and the equivalent width stalls. This is not a limitation for \({}^{13}\mathrm{CO}\) and \(\mathrm{C}{}^{18}\mathrm{O}\), whose abundances are $\approx$2 orders of magnitude lower than \({}^{12}\mathrm{CO}\) and therefore remain in the weak-line regime. In contrast, the \Kband{} contains many intrinsically weaker, unsaturated \({}^{12}\mathrm{CO}\) lines that constrain the carbon abundance. Together with the dense H$_2$O line forest that constrains oxygen, the \Kband{} therefore supplies the elemental C and O abundances that the saturated \Mband{} \({}^{12}\mathrm{CO}\) lines cannot.

The \Mband{} posterior for C/O shows a moderate degeneracy with the isotopic ratios (Fig.~\ref{fig:cornerplot}) because of that saturation. Combining the two bands breaks the degeneracy: the unsaturated \Kband{} \({}^{12}\mathrm{CO}\) lines fix the carbon abundance, and the joint fit then yields a stable C/O and isotopic ratios.
Independent free-$P$--$T$ fits of the two bands alone give discrepant C/O values ($\approx$0.44 for \Mband{} versus $\approx$0.59 for \Kband{}), but repeating the \Mband{} retrieval with the $P$--$T$ profile fixed to the best-fit \Kband{} (SPIRou) solution raises the \Mband{} C/O to $0.55\pm0.03$.
The apparent band-to-band C/O offset is therefore an artifact of the distinct preferred-temperature structures rather than a fundamental inconsistency between the two CO band systems (Sec.~\ref{sec:teff}).

\subsection{Elemental abundances and planet formation}

The retrieved C/O ratio of $0.60^{+0.02}_{-0.02}$ provides a stellar reference point for interpreting the formation
and evolution of the TRAPPIST-1 planets.
This C/O ratio and the metallicity of $[\mathrm{M/H}] = 0.00^{+0.06}_{-0.06}$ are consistent with prior characterizations of TRAPPIST-1 as having roughly solar composition \citep{gillon:2017,davoudiUpdatedSpectralCharacteristics2024}.
The TRAPPIST-1 planets may have migrated into their current location from the outer regions of the disk, beyond the water snowline (e.g.\ \citealt{ormelFormationTRAPPIST1Other2017}). In such a scenario, the elemental C/O ratios of the planets may differ from the stellar value through gas and solid accretion from chemically distinct disk regions \citep{mordasini:2016,espinoza:2017,ormelFormationTRAPPIST1Other2017}. Our measurement establishes the stellar benchmark but does not by itself constrain individual planetary C/O ratios or formation pathways. Such comparisons require atmospheric measurements of the planets themselves \citep{ormelFormationTRAPPIST1Other2017,schneebergerFormationTrappist1System2024}.

\subsection{Isotopic ratios as tracers of Galactic chemical evolution}
\label{sec:isotopic}

The carbon and oxygen isotopic ratios reflect chemical processing in the material from which TRAPPIST-1 formed. Our value $\cisoratio = 154^{+17}_{-16}$ exceeds the solar-system ratio of 89 \citep{lyons:2018} and follows the GCE trend in Fig.~\ref{fig:isotope_ratios}, but it is not an independent age indicator. Chemical-evolution models predict rising \({}^{13}\mathrm{C}\) over time from CNO-cycle processing in intermediate-mass stars \citep{prantzos:2018,romano:2022}. Higher \cisoratio{} ($>120$) is common in older objects, including GJ~745~AB \citep[old nearby M dwarf binary with $\cisoratio=200$--300;][]{crossfield:2019a}, Barnard's Star (7--10\,Gyr, $\cisoratio>170$, \citealt{picos:2025}), 3I/ATLAS (10--12\,Gyr, $\cisoratio=144$--196, \citealt{cordinerJWSTSpectroscopy3I2026}), and HD~19467~B (8--10\,Gyr, $\cisoratio=151^{+17}_{-16}$, \citealt{gonzalezpicosJWSTHighcontrastSpectroscopy2026a}). That trend matches the TRAPPIST-1 age of \(7.6 \pm 2.2\)\,Gyr \citep{burgasserAgeTRAPPIST1System2017}, although work on low-resolution spectroscopy still allows for a wider range due to mixed field and youth spectral features \citep{gonzalesReanalysisFundamentalParameters2019}. A caveat is that low-metallicity, massive, fast rotators can produce primary $^{13}$C early in Galactic history and help explain the low $\cisoratio$ of halo stars \citep[e.g., $\cisoratio = 33^{+12}_{-6}$ for HD\,140283][]{spite2021}. Hot CNO burning in nova outbursts also synthesizes $^{13}$C; the rarity of those events and the sensitivity of the yield to the progenitor configuration can add scatter in $\cisoratio$ at [M/H]~$> -1$ \citep[see][and references therein]{romano2017}.

\begin{figure}
    \centering
    \includegraphics[width=\columnwidth]{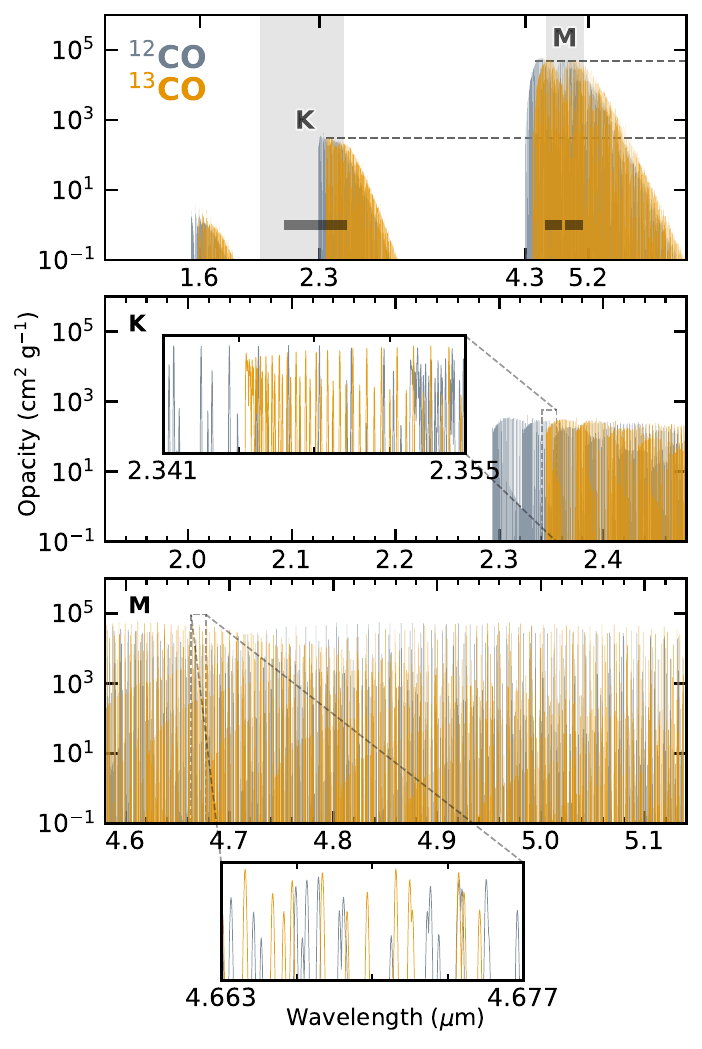}
    \caption{Opacity spectra of \({}^{12}\mathrm{CO}\), \({}^{13}\mathrm{CO}\), and \(\mathrm{C}{}^{18}\mathrm{O}\) computed with \texttt{pyROX} \citep{regtPyROXRapidOpacity2025} from the line lists of \citealt{liROVIBRATIONALLINELISTS2015} at \(T=2500\) K, $P=1$ bar, and solar C/O, broadened to \(\mathcal{R}=100{,}000\). Opacity is in cm$^2$\,g$^{-1}$ (per unit mass). \textit{Top:} full IR range. SPIRou and Keck/NIRSPEC coverage are indicated by shaded boxes and the typical ground-based \Kband{} and \Mband{} coverage as gray bands. \textit{Middle:} \Kband{} first-overtone (\(\nu=2\rightarrow 0\)) bands. \textit{Bottom:} \Mband{} fundamental (\(\nu=1\rightarrow 0\)) bands. \Mband{} opacity is $\sim$100 times higher than in the \Kband{}.}
    \label{fig:CO_opacity}
\end{figure}

Our oxygen isotopic ratio, $\oisoratio = 490^{+78}_{-64}$, agrees with other solar-metallicity M dwarfs (e.g., GJ~849, $\oisoratio = 490^{+171}_{-101}$, \citealt{picos:2025}) and is more precise despite the relative faintness of TRAPPIST-1. It also agrees with the solar value ($525\pm21$, \citealt{lyons:2018}) and the present-day ISM ($557 \pm 30$, \citealt{wilsonIsotopesInterstellarMedium1999}) within uncertainties. \({}^{18}\mathrm{O}\) is produced in massive stars, and uncertain yields from fast-rotating, metal-poor progenitors limit \oisoratio{} as an age indicator \citep{limongiPresupernovaEvolutionExplosive2018,prantzosChemicalEvolutionRotating2018,romanoEvolutionCNOIsotopes2019}. Access to the fundamental CO band in the \Mband{} is therefore valuable (Fig.~\ref{fig:CO_opacity}): detecting \(\mathrm{C}{}^{18}\mathrm{O}\) in the \Kband{} remains difficult even for bright M dwarfs, because the CO opacity is roughly 100 times lower than in the \Mband{} (Fig.~\ref{fig:CO_opacity}) and water opacity partially blankets the CO lines.

Water isotopologues can also constrain oxygen isotope ratios \citep[e.g.][]{regtESOSupJupSurvey2026}, but H$_2$O dissociates at lower temperatures than CO and is largely absent from the gas phase in mid- to early-M atmospheres ($T \gtrsim 3500$~K). At TRAPPIST-1 temperatures water is abundant and should in principle yield an isotopic measurement. In practice we recover only a weak \(\mathrm{H}_2^{18}\mathrm{O}\) signal and unreliable water-based ratios, consistent with systematic wavelength offsets in some water lines reported by \citet{picos:2025}. Water-based constraints are further limited by incomplete or inaccurate high-temperature line lists. Molecular databases have improved over the past decade \citep{rothmanHITEMPHightemperatureMolecular2010,polyanskyExoMolMolecularLine2018,tennyson2024ReleaseExoMol2024}, but the 2000--4000~K regime still needs care. Significant line mismatches have been found in the near-infrared spectrum of a mid-M dwarf observed with SPIRou, with the \Kband{} slightly less affected than bluer wavelengths \citep[e.g.][]{jahandarComprehensiveHighresolutionChemical2024}. The CO isotopologues are better constrained in our data and give the more reliable isotopic ratios, as expected from the greater stability and fewer transitions of CO relative to H$_2$O \citep{liROVIBRATIONALLINELISTS2015}.

We also attempted to retrieve C$^{17}$O but find no evidence for it in our data. Because the posterior piles up at the upper prior bound, we quote a $\sim$99\% lower limit of $\log_{10}({}^{12}\mathrm{C}^{16}\mathrm{O}/{}^{12}\mathrm{C}^{17}\mathrm{O})\geq3.36$, corresponding to $\ratio{O}{16}{17} > 2300$ \citep[cf.\ $\ratio{O}{16}{17} = 2738 \pm 118$ for the Sun;][]{ayresSUNLIGHTEREARTH2013} or $\ratio{O}{18}{17} > 4.3$ \citep[cf.\ $\ratio{O}{18}{17} \approx 4$ for giant stars in the Milky Way;][and references therein]{romano:2022}. A direct measurement of \ratio{O}{18}{17}{} in unevolved stars will likely require next-generation facilities (e.g., ELT/METIS, \citealt{brandlMETISMidinfraredELT2021}).

\begin{figure*}
    \includegraphics[width=\textwidth]{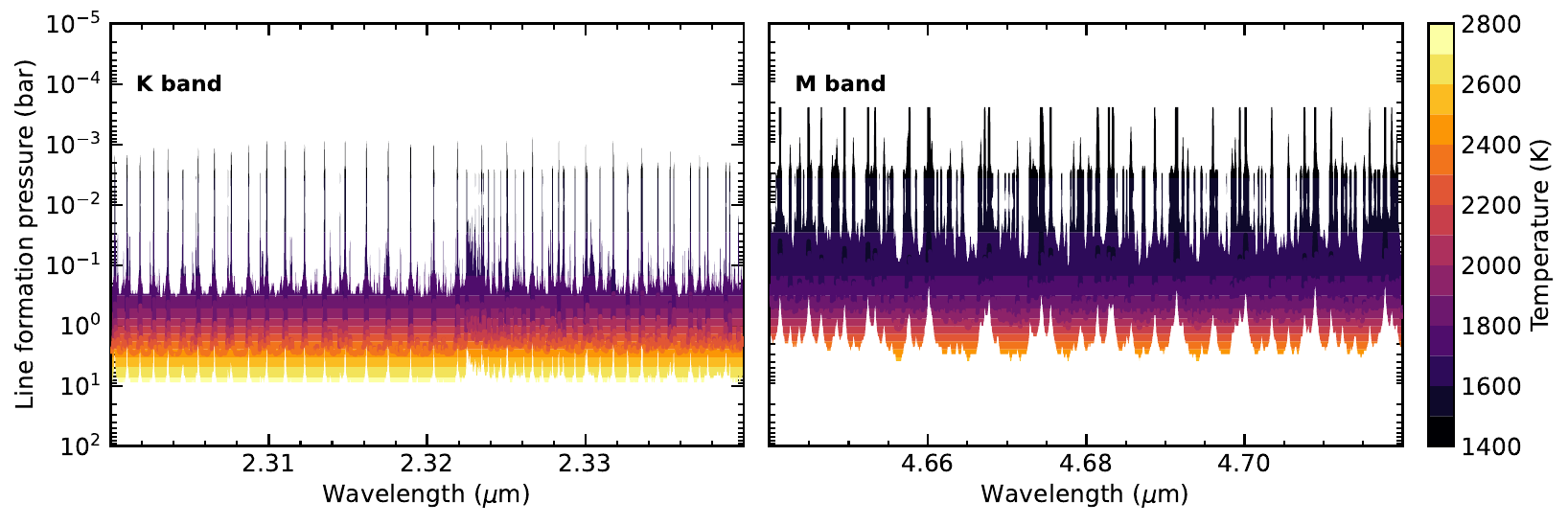}
    \caption{Line-formation temperature versus wavelength for the joint best-fit retrieval, computed as the pressure-weighted temperature where the emission-contribution function peaks. The \Kband{} probes deeper (hotter) layers than the \Mband{}.}
    \label{fig:line_formation_temperature_wavelength}
\end{figure*}

Oxygen isotope ratios in the solar system have been used to constrain the natal environment and planet-building material \citep{lyonsCOSelfshieldingOrigin2005a,gaidosOxygenIsotopicComposition2009,nittlerGalacticChemicalEvolution2012}, but extending such work to field stars is difficult because optical spectra (typically used to characterize hotter stars) lack strong O-bearing features. For solar-type and hotter stars, CO weakens or dissociates, which makes CO isotopologues harder to measure \citep[e.g.,][]{coria:2023,tsujiMolecularAbundanceStellar1964}. Cool dwarfs such as K- and M-type stars exhibit strong CO absorption lines that enable the detection of CO isotopologues and inference of carbon and oxygen isotopic ratios.

\begin{figure}[hb]
    \centering
    \includegraphics[width=\columnwidth]{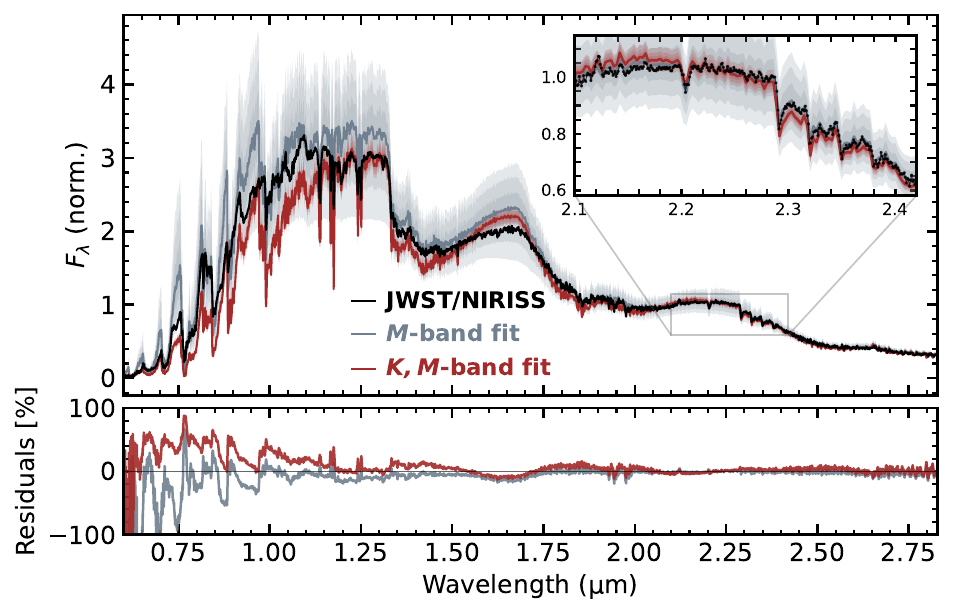}
    \caption{Low-resolution comparison to JWST/NIRISS \citep{limAtmosphericReconnaissanceTRAPPIST12023}. The NIRISS spectrum was not included in the retrieval or fitted separately; it is overlaid here for reference. Observed spectrum (black, calibrated flux density). Joint \Kband+\Mband{} retrieval (brown) and \Mband-only retrieval (gray) with 68\% and 95\% posterior envelopes propagated from the high-resolution fit. \textit{Inset:} SPIRou \Kband{} region used to set the relative normalization between model and data.}
    \label{fig:bestfit_spectrum_niriss}
\end{figure}

\subsection{Atmospheric temperature structure}
\label{sec:teff}
High-resolution spectroscopy provides strong constraints on stellar atmospheric pressure--temperature profiles, but these constraints come mainly from lines in the spectrum rather than the overall spectral energy distribution. Our line-based effective temperature is 200--300~K below SED-based values. The joint \Kband+\Mband{} retrieval yields $T_{\mathrm{eff}} = 2325 \pm 13$~K, compared to $T_{\mathrm{eff}} = 2612^{+57}_{-56}$~K for the \Mband-only retrieval and \(T_{\mathrm{eff}}(\mathrm{SED}) = 2569 \pm 28\) K \citep{davoudiUpdatedSpectralCharacteristics2024}. We retrieve distinct temperature structures when fitting the \Mband{} and \Kband{} spectra independently, and the joint fit is dominated by the \Kband{} given its larger number of data points and higher S/N. In the joint fit we infer a cooler deep atmosphere than both the self-consistent SPHINX-II models \citep{iyerSPHINXDwarfSpectral2025} and our \Mband-only fit. The regions of the atmosphere probed by the \Kband{} and \Mband{} are shown in Fig.~\ref{fig:line_formation_temperature_wavelength}, indicating that the \Kband{} probes deeper in the atmosphere than the \Mband{}. At the bottom of the K-band photospheric region, the temperatures approach 2800~K, while the M-band probes layers near 2400~K.
Under single-column assumptions this band-to-band $P$--$T$ tension persists: a free \Mband{} profile fits the NIRSPEC data substantially better than one fixed to the SPIRou solution (strong preference from Bayesian model comparison), yet that same free \Mband{} profile fares poorly when extended to shorter wavelengths. Each band can be fit well in isolation, but a single pressure--temperature structure struggles to simultaneously describe both wavelength regimes; the joint retrieval is likely a compromise between them, suggesting that more complex atmospheric models will ultimately be needed to fit multiple wavelength ranges of TRAPPIST-1 simultaneously.

As a check against the SED, we compute low-resolution spectra (\(\mathcal{R}=800\)) and compare them to JWST/NIRISS (Fig.~\ref{fig:bestfit_spectrum_niriss}, data from \citealt{limAtmosphericReconnaissanceTRAPPIST12023,davoudiUpdatedSpectralCharacteristics2024}). Both models match the overall continuum shape and the reddest region (\(>2.0\,\mu\mathrm{m}\)), but overestimate the flux at the H-band peak (1.55--1.65\,\(\mu\mathrm{m}\)) and at shorter wavelengths, in line with the reported mix of field and youth spectral features \citep{gonzalesReanalysisFundamentalParameters2019}. The joint \Kband+\Mband{} solution also predicts a lower blue continuum, as expected from its lower \(T_{\mathrm{eff}}\).

We interpret our retrieved \(T_{\mathrm{eff}}\) as the temperature structure that best reproduces the high-resolution line spectrum, rather than a standalone estimate of the global photospheric \(T_{\mathrm{eff}}\). We derive bolometric \(T_{\mathrm{eff}}\) by integrating the best-fit model flux from 0.3 to 20\,\micron\ after renormalizing the high-resolution spectrum to the SPIRou continuum level; the quoted uncertainty reflects the posterior on the atmospheric parameters only and not absolute flux calibration. Full-SED constraints remain better suited to the global photospheric temperature \citep{gonzalesReanalysisFundamentalParameters2019,davoudiUpdatedSpectralCharacteristics2024}. The offset between our line- and SED-based temperatures is plausibly linked to surface heterogeneity, as in spot--faculae models invoked for JWST transmission spectra of TRAPPIST-1~b \citep{rackhamTransitLightSource2018,limAtmosphericReconnaissanceTRAPPIST12023}. \citet{rathckeStellarContaminationCorrection2025} find that TRAPPIST-1 is well described by a two-temperature photosphere with warm ($\approx$2600~K) and cool ($\approx$2000~K) components each covering roughly half of the visible surface; from their best-fit model they report $T_{\mathrm{eff}} \approx 2324$~K, in agreement with our line-based retrieval.

Our retrieved C/O is less affected by this \(T_{\mathrm{eff}}\) tension. It is anchored by unsaturated \Kband{} \({}^{12}\mathrm{CO}\) lines and \Mband{} \({}^{13}\mathrm{CO}\) and \(\mathrm{C}{}^{18}\mathrm{O}\) features (Sec.~\ref{sec:results}). Missing or inaccurate opacity sources in the model could contribute to the mismatch, particularly in the 2000--3000\,K regime relevant to TRAPPIST-1, which remains largely unexplored for high-resolution spectroscopy. We calculated the contribution of potential missing continuum opacity sources such as He$^{-}$ free-free and H$_2^{-}$ continua but found them to be negligible \citep{grayObservationAnalysisStellar2022}.

\subsection{Two-column model}
\label{sec:two_column}

Motivated by the offset between our line- and SED-based temperatures, we perform an \emph{exploratory}, prior-driven test of whether surface heterogeneity can reconcile these values by implementing a two-column photosphere with warm and cool components, broadly corresponding to the quiet photosphere and spots.
Each component is assigned a separate pressure--temperature profile interpolated from the SPHINX-II grid \citep{iyerSPHINXDwarfSpectral2025}, with chemical-equilibrium composition and emergent flux computed using the same radiative-transfer setup as in Sec.~\ref{sec:analysis}.
The disk-integrated spectrum is the covering-fraction-weighted sum
\begin{equation}
  F = \left(1 - f_{\mathrm{cool}}\right)\, F_{\mathrm{warm}} + f_{\mathrm{cool}}\, F_{\mathrm{cool}},
  \label{eq:two_column}
\end{equation}
where \(f_{\mathrm{cool}}\) is the fractional surface coverage of the cool component.
Technical details of the implementation, including the informative priors that enforce a non-zero cool contribution, and the corresponding fits and posterior comparisons are given in Appendix~\ref{app:two_column_residuals}.

We compare this two-column SPHINX model to a one-column SPHINX retrieval and to the flexible one-column retrieval of Sec.~\ref{sec:analysis}.
Model comparison via the log Bayes factor strongly favors the flexible one-column solution over both SPHINX parameterizations, with \(\ln B > 100\) in each case \citep{thorngrenBayesianModelComparison2026}.
The three retrievals yield comparable \Kband{} fits but substantially different bolometric effective temperatures (\(T_{\mathrm{eff}}\approx 2300\)--$2550$~K), while C/O remains stable across the tested parameterizations.
The statistical preference for the flexible model is driven mainly by the \Mband{} fit quality (Appendix~\ref{app:two_column_residuals}).
For the tested wavelength range and S/N, high-resolution spectroscopy alone does not distinguish one- and two-column solutions with similar \Kband{} fits but different bolometric \(T_{\mathrm{eff}}\); complementary photometric or time-resolved constraints will be needed to refine the surface coverage and temperature structure.

\subsection{Radial and rotational velocity}

From the high-resolution \Kband{} SPIRou spectrum we measure a systemic radial velocity of $v_{\mathrm{rad}} = -52.97 \pm 0.04$~km\,s$^{-1}$, corrected to the barycenter at MJD~59065. We allow separate radial velocities for each NIRSPEC order because the manual reduction leaves order-dependent wavelength zero-point offsets that are not fully tied to the SPIRou solution; the fitted values of $-3.4$ and $-78.0$~km\,s$^{-1}$ for orders 16 and 15 are nuisance parameters that absorb these offsets rather than independent astrophysical velocities. Relative line positions \emph{within} each order still constrain the isotopologues, and the cross-correlation analysis (Fig.~\ref{fig:cross_correlation}) shows $^{13}$CO and C$^{18}$O peaks at the velocities expected after applying the order-specific shifts. The cubic wavelength solutions in Appendix~\ref{app:nirspec} reproduce telluric and stellar lines in the B9V calibrator with rms scatters of 0.014 and 0.029\,nm ($\sim$1--2~km\,s$^{-1}$ at 4.7\,\micron), much smaller than the fitted order offsets, indicating that the latter primarily reflect imperfections in the telluric division and stacking rather than failures of the wavelength polynomial itself.

Rotational constraints on TRAPPIST-1 favor a slowly rotating star. Analysis of K2 photometry reported $P_{\mathrm{rot}} \approx 3.3$~d \citep{luger:2017}, which, for the measured radius \citep{grootelStellarParametersTrappist12018}, implies $v_{\mathrm{eq}} \sim 2$~km\,s$^{-1}$. An early optical estimate of $v\sin i = 6 \pm 2$~km\,s$^{-1}$ \citep{reinersBasriMagneticTopology2010} was hard to reconcile with that period, and \citet{roettenbacherStellarActivityTRAPPIST12017} cautioned that evolving activity can bias both photometric and spectroscopic rotation indicators. Later high-resolution work has instead placed $v\sin i$ near 2~km\,s$^{-1}$, including CARMENES \citep{reinersCARMENESSearchExoplanets2018}, Subaru/IRD Rossiter--McLaughlin spectroscopy (1.49$^{+0.36}_{-0.37}$ km\,s$^{-1}$, \citealt{hiranoSpinOrbitAlignmentTRAPPIST12020}), and MAROON-X (2.1 $\pm$ 0.3 km\,s$^{-1}$, \citealt{bradyObliquitiesTRAPPIST1Planets2024}).

Our joint SPIRou+Keck retrieval yields $v\sin i = 2.15 \pm 0.19$~km\,s$^{-1}$, confirming the slow rotation inferred from photometry and prior spectroscopy. This value agrees well with the MAROON-X measurement of \citet{bradyObliquitiesTRAPPIST1Planets2024}, but should be interpreted cautiously. At $\mathcal{R} \approx 70{,}000$, SPIRou resolves only $\sim$4.3~km\,s$^{-1}$ per spectral element and is therefore near the sensitivity limit for rotational broadening at this level. Higher-resolution observations, for example with CRIRES+ at the Very Large Telescope ($\mathcal{R}=100{,}000$, \citealt{dornCRIRESSkyESO2023}) or PEPSI at the Large Binocular Telescope \citep[$\mathcal{R}=300{,}000$]{strassmeier:2008}, could provide a superior constraint.

\section{Conclusions}
\label{sec:conclusions}

We have derived the first stellar C/O ratio and the first carbon- and oxygen-isotope ratios for TRAPPIST-1 using high-resolution ground-based spectroscopy at \Mband{} (Fig.~\ref{fig:mband}) and \Kband{} (Fig.~\ref{fig:bestfit_spec_spirou}). Under our preferred flexible one-column retrieval, we obtain $[\mathrm{M/H}]=0.00^{+0.06}_{-0.06}$, C/O $=0.60^{+0.02}_{-0.02}$, $\cisoratio = 154^{+17}_{-16}$, and $\oisoratio = 490^{+78}_{-64}$. These quoted intervals are statistical posterior credible intervals (16th--84th percentiles). Both $^{13}$CO and C$^{18}$O are detected in the Keck/NIRSPEC \Mband{} spectra (Fig.~\ref{fig:cross_correlation}), while the CFHT/SPIRou \Kband{} spectra most tightly constrain the bulk elemental abundances (Fig.~\ref{fig:cornerplot}). We do not detect $^{17}$O; a lower limit of $\ratio{O}{16}{17}>2300$ is consistent with solar and giant-star values.

An exploratory, prior-driven two-column retrieval (Sec.~\ref{sec:two_column}) shows that comparable \Kband{} fits can yield bolometric $T_{\mathrm{eff}}$ values spanning $\sim$2300--2550~K depending on the assumed photosphere coverage and temperature contrast, while C/O remains stable. For the tested wavelength range and S/N, high-resolution data alone do not distinguish this scenario from a free-$P$--$T$ single-column solution. We present the joint \Kband+\Mband{} single-column retrieval as our best estimate of the atmospheric structure and composition, while noting that the remaining band-to-band $P$--$T$ discrepancy might be resolved with more complex models (e.g., two-column treatments) or more representative physical temperature structures.

Despite hosting the only known system of seven terrestrial planets, TRAPPIST-1 is chemically a relatively ordinary star. Fig.~\ref{fig:isotope_ratios} places TRAPPIST-1 among the Sun, other cool dwarfs, and GCE models \citep{romano:2022}. Its $\oisoratio$ matches that population, its C/O matches bright solar-neighborhood stars \citep{fortney:2012}, and its $\cisoratio$ sits somewhat above the Sun and some nearby dwarfs at similar metallicity. More homogeneous measurements of this population would help map the scatter. The full population of cool dwarfs with measured $\cisoratio$ and $\oisoratio$ plotted in Fig.~\ref{fig:isotope_ratios} also spans only a single order of magnitude in [M/H]; GCE models adopting different nucleosynthesis prescriptions increasingly diverge at lower metallicities \citep{romano2017,romanoEvolutionCNOIsotopes2019}, indicating that such measurements in stars with [M/H]$\lesssim$-0.5 would help constrain free parameters of stellar evolution and nucleosynthesis models.

Deep ground-based \Mband{} spectroscopy remains the practical route to $^{18}$O in faint M dwarfs \citep[cf.][]{crossfield:2019a,coria:2023}, while \Kband{} data give the cleanest access to elemental C and O and, in amenable targets, to $^{13}$C at lower sensitivity. If planetary atmospheres are detected in this system, these stellar isotope ratios will be the natural comparison.

Finally, our observations also show the power of deep ground-based \Mband{} spectroscopy to characterize faint, red objects. \Mband{} observations of the directly imaged planet $\beta$~Pic~b revealed CO and H$_2$O at S/N$=$6--7 in just 2.4\,hr of VLT/CRIRES+ \Mband{} observations \citep{parker:2024}. That planet has $m_M\approx 10.9$ \citep{morzinski:2014}, roughly one magnitude fainter than TRAPPIST-1 ($m_{W2}=9.9$). The ELT/METIS instrument \citep{brandlMETISMidinfraredELT2021} will offer substantially higher sensitivity than the VLT, with $\sim$23$\times$ greater collecting area and reduced backgrounds due to the smaller diffraction-limited PSF; together with our TRAPPIST-1 results and those for $\beta$~Pic~b, this suggests that METIS and similar instruments will extend isotopic measurements to many more exoplanets, brown dwarfs, and cool dwarf stars.

\begin{acknowledgments}

We would like to thank the anonymous referee for the constructive comments. We thank our funding sources: I.J.M.C.\ acknowledges support from NASA/Keck via JPL (program 2020B\_N102), from NASA ICAR grant 80NSSC21K0597, and from NSF AAG grant 2108686. D.G.P., I.S., and S.d.R.\ acknowledge NWO grant OCENW.M.21.010. D.R.\ acknowledges support from INAF through program Finanziamento della Ricerca Fondamentale, Theory Grant ``An in-depth theoretical study of CNO element evolution in galaxies'', Fu.~Ob.~1.05.12.06.08. E.G.\ acknowledges support from NASA Exoplanets Research Program award 80NSSC20K0957. This work used the Dutch national e-infrastructure with the support of the SURF Cooperative using grant no.\ EINF-4556.

The authors wish to recognize and acknowledge the very significant cultural role and reverence that the summit of Maunakea has always had within the Native Hawaiian community. We are most fortunate to have the opportunity to conduct observations from this mountain.

Some of the data presented herein were obtained at Keck Observatory, which is a private 501(c)3 non-profit organization operated as a scientific partnership among the California Institute of Technology, the University of California, and the National Aeronautics and Space Administration. The Observatory was made possible by the generous financial support of the W. M. Keck Foundation.

Based on observations obtained at the Canada--France--Hawai'i Telescope (CFHT) which is operated by the National Research Council of Canada, the Institut National des Sciences de l'Univers of the Centre National de la Recherche Scientifique of France, and the University of Hawai'i. CFHT is located on Maunakea on Hawai'i Island, a mountain of considerable cultural, natural, and ecological significance.

This research has made use of the Keck Observatory Archive (KOA), which is operated by the W. M. Keck Observatory and the NASA Exoplanet Science Institute (NExScI), under contract with the National Aeronautics and Space Administration.

We thank X.\ Bonfils and the SPIRou team of run 20BF18, whose public data we used.

We thank T.\ Greene for useful discussions about the Keck/NIRSPEC observations.

\end{acknowledgments}

\begin{contribution}

Author contributions to paper:  DGP downloaded and combined the SPIRou data, and conducted the various spectroscopic analyses.  IC, JL, EG, and EM proposed for the \Mband{} data, IC and DC conducted the observations, and IC reduced the data and extracted the spectrum.   DGP and IC led the paper write-up, with all other authors also contributing.

\end{contribution}

\facilities{Keck:II (NIRSPEC), CFHT (SPIRou), JWST (NIRISS)}

\software{Python 3.11,
          petitRADTRANS \citep{mollierePetitRADTRANSPythonRadiative2019},
          pyROX \citep{regtPyROXRapidOpacity2025},
          FastChem \citep{kitzmannFastchemCondEquilibrium2023},
          PyMultiNest \citep{buchnerPyMultiNestPythonInterface2016},
          MultiNest \citep{ferozMultiNestEfficientRobust2009},
          APERO \citep{cookAPEROPipelinEReduce2022}}

\bibliography{ms}{}
\bibliographystyle{aasjournalv7}

\appendix

\section{NIRSPEC Data Reduction}
\label{app:nirspec}
We manually reduced the NIRSPEC \Mband{} data using a set of custom Python routines and spectral analysis routines, adapting some of the tools used in previous studies \citep{crossfield:2011,crossfield:2013b}.  We stacked a set of flat frames taken with an internal calibration lamp, then traced and continuum-normalized each order to generate a master flat-field frame for each night. We divided each frame by this flat field, and constructed a single stacked frame for each target during each night: ``A'' position frames were added to the stack, and ``B'' position frames were subtracted. This removes most sky and instrumental backgrounds, and results in two spectral traces with sufficient S/N for subsequent extraction.  We fit a polynomial to each trace's position across each order, and extracted a 60$\times$2048 pixel swath of detector pixels centered on the trace.  To account for the slight tilt of residual emission features \citep[due to NIRSPEC's quasi-Littrow configuration;][]{mclean:1998} we interpolated each of the 60 rows to de-tilt each order.
At each wavelength, we then subtracted a median background level and summed the remaining (stellar) flux in the central 6 pixels.

We manually solved for each order's wavelength solution by identifying matching lines in the spectra of HR~8865 and in a high-resolution telluric atlas. In both cases a cubic wavelength solution gave the optimal fit; we achieved a final rms of 0.014\,nm and 0.029\,nm for orders 15 and 16, respectively. The final wavelength solutions, in \micron\ and in terms of pixel number $x$ (from 0--2047 inclusive), are:
\begin{align}
  \lambda_{15} &= 2.76403198\mathrm{e-}13 x^3 + 5.9601135\mathrm{e-}10 x^2 - 5.489685\mathrm{e-}05 x +  5.04101 \\
  \lambda_{16}  &= 2.12594653\mathrm{e-}13 x^3 + 6.5459974\mathrm{e-}10 x^2 - 5.141004\mathrm{e-}05 x+  4.72835
\end{align}

After aligning the TRAPPIST-1 and HR~8865 spectra for each nod position and each night using telluric absorption features, we divided the TRAPPIST-1 spectra by that of HR~8865 to remove the effects of moderate telluric absorption \citep{vacca:2003}.  We then aligned the six (two nods times three nights) resulting telluric-corrected spectra, continuum-normalized them,  and stacked them to construct a final stellar spectrum $f_\lambda$. We estimate the uncertainty on the spectrum by calculating the standard error of the mean ($\sigma_{f_\lambda} = \sigma/\sqrt{6}$) in each wavelength channel, smoothing the resulting uncertainty spectrum with a 5-pixel-width median filter, and enforcing a minimum uncertainty of $\sigma_{f_\lambda}\geq0.05$ in normalized flux units (equivalent to capping S/N at 20). Bad pixels were masked interactively during extraction; no separate cosmic-ray rejection was applied beyond the nod-subtraction procedure.

\section{Machine-readable spectrum}
\label{app:spectrum_table}

Table~\ref{tab:trappist1_spectrum} lists a sample of the continuum-normalized \Kband{} and \Mband{} spectra together with the joint best-fit model.
$F_\lambda$, $\sigma_{F_\lambda}$, and Model are continuum-normalized fluxes.
Orders 0--4 are SPIRou \Kband{}; orders 5--6 are Keck/NIRSPEC \Mband{}.
Adjacent SPIRou orders can overlap in wavelength; select on \texttt{Order} and sort by \texttt{Index} before interpreting the spectrum.
Missing values in the machine-readable file are written as NaN.
The full table is provided as the machine-readable file \texttt{trappist1\_spectrum\_spirou\_nirspec.txt}.

\startlongtable
\begin{deluxetable}{c@{\hspace{2em}}c@{\hspace{2em}}c@{\hspace{2em}}c@{\hspace{2em}}c@{\hspace{2em}}c}
\tablecaption{Sample of the continuum-normalized spectrum and best-fit model for TRAPPIST-1.\label{tab:trappist1_spectrum}}
\tablehead{
\colhead{$\lambda$ [\micron]} &
\colhead{$F_\lambda$} &
\colhead{$\sigma_{F_\lambda}$} &
\colhead{Model} &
\colhead{Order} &
\colhead{Index}
}
\startdata
2.100593 & 0.893 & 0.018 & 1.022 & 0 & 0 \\
2.100611 & 0.917 & 0.016 & 1.048 & 0 & 1 \\
\nodata & \nodata & \nodata & \nodata & \nodata & \nodata \\
4.630033 & 0.683 & 0.082 & 0.772 & 5 & 30 \\
4.630079 & 0.781 & 0.064 & 0.732 & 5 & 31 \\
\nodata & \nodata & \nodata & \nodata & \nodata & \nodata \\
\enddata
\end{deluxetable}

\section{Continuum fitting with spline-based spectral decomposition}
\label{app:spline_continuum}

High-resolution spectra are often continuum-normalized before analysis, but classical high-pass filters, sliding-window medians, and low-order polynomial divisions can introduce systematic biases and are sensitive to bad pixels and telluric residuals \citep[e.g.,][]{nessCannonDatadrivenApproach2015}. Joint forward modeling of atmospheric templates and flexible continua avoids the biases of prenormalization and has recently been applied to spectroscopy of M dwarfs and directly imaged companions \citep{ruffioJWSTTSTHighContrast2023,gonzalezpicosESOSupJupSurvey2025,picos:2025,ruffioJupiterlikeUniformMetal2026}.
Following \citet{ruffioDetectingExomoonsRadial2023}, we write the continuum as a \emph{linear} combination of spline modes.
Although the continuum shape is a nonlinear function of wavelength, it is linear in the amplitudes $\boldsymbol{\phi}$, which permits efficient weighted least-squares fitting at each likelihood evaluation.
Our implementation reproduces the algorithmic core of that approach and is adapted from the \texttt{breads} software package\footnote{\url{https://github.com/jruffio/breads}}.
We use $N_k=35$ knots per spectral segment by default and verify that the recovered C/O and isotope ratios are insensitive to $N_k$.

Let $\{x_i\}_{i=0}^{N-1}$ denote pixel indices and $\{u_j\}_{j=0}^{N_k-1}$ the uniformly spaced knot locations ($N_k$ knots).
We construct cubic B-spline basis functions $B_j^{(k)}(x)$ of degree $k$ (default $k=3$, as in \citealt{ruffioDetectingExomoonsRadial2023}), which add up to unity $\sum_j B_j^{(k)}(x)=1$ and are smoothly connected.
Given a template spectrum $s_i$ (for example, a synthetic model resampled onto the data grid), the $N_k$ continuum modes are
\begin{equation}
  M_{ji} = B_j^{(k)}(x_i)\, s_i,
  \label{eq:spline_modes}
\end{equation}
and the continuum-modulated model at each pixel is
\begin{equation}
  c(x_i) = \sum_{j=0}^{N_k-1} \phi_j\, M_{ji}.
  \label{eq:continuum_model}
\end{equation}
Knots are placed uniformly in pixel space, extended slightly beyond the spectral domain to suppress edge artifacts:
\begin{equation}
  u_j = -1 + j\,\frac{N + 1}{N_k - 1}, \qquad j = 0, 1, \ldots, N_k - 1.
  \label{eq:knot_placement}
\end{equation}
Basis functions are evaluated only for pixels interior to the knot interval ($u_0 < x_i < u_{N_k-1}$).
The number of knots $N_k$ and spline degree $k$ control the continuum flexibility; we require $N_k \ge 2$ and $k \le N_k - 1$.

For observed flux $\mathbf{d}$ with diagonal inverse variance $\mathbf{C}^{-1}$, the weighted least-squares estimate of $\boldsymbol{\phi}$ minimizes
\begin{equation}
  \chi^2(\boldsymbol{\phi}) =
  \left(\mathbf{d} - \mathbf{M}^{\mathsf T}\boldsymbol{\phi}\right)^{\mathsf T}
  \mathbf{C}^{-1}
  \left(\mathbf{d} - \mathbf{M}^{\mathsf T}\boldsymbol{\phi}\right),
  \label{eq:spline_chi2}
\end{equation}
which we solve with a non-negative least-squares algorithm rather than direct matrix inversion for numerical stability \citep{lawsonSolvingLeastSquares1995}. The optimized amplitudes define a profile likelihood at each nested-sampling step; we do not analytically marginalize over $\boldsymbol{\phi}$ during nested sampling.

\begin{figure}[t]
    \centering
    \includegraphics[width=\textwidth]{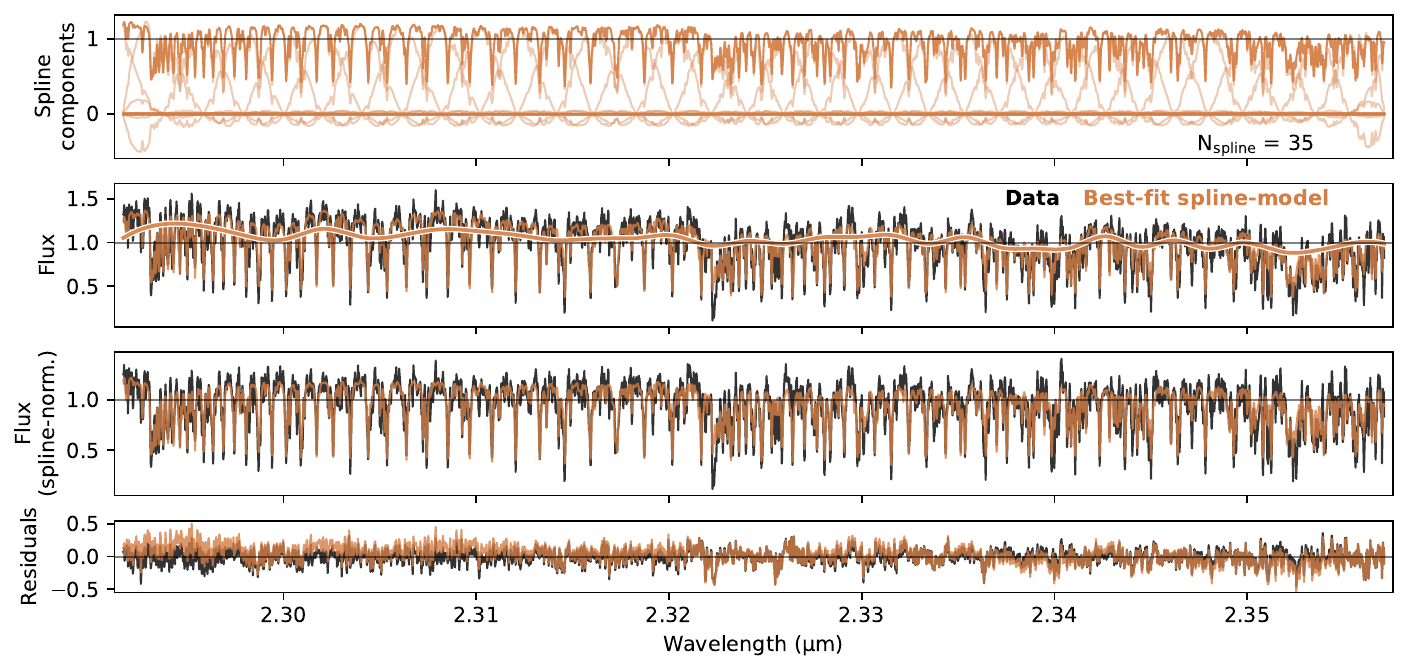}
    \caption{Spline-based continuum normalization applied to a SPIRou \Kband{} order. \textit{Top:} individual B-spline basis functions ($N_k=35$). \textit{Second:} observed spectrum (black) with the best-fit multiplicative spline envelope (orange). \textit{Third:} data and spline model after division by the fitted continuum. \textit{Bottom:} fit residuals without (orange) and with (black) the spline continuum included in the forward model; including the spline greatly reduces systematic continuum structure in the residuals.}
    \label{fig:spline_continuum}
\end{figure}

\section{Two-column model comparison}
\label{app:two_column_residuals}

\begin{figure}
    \centering
    \includegraphics[width=\textwidth]{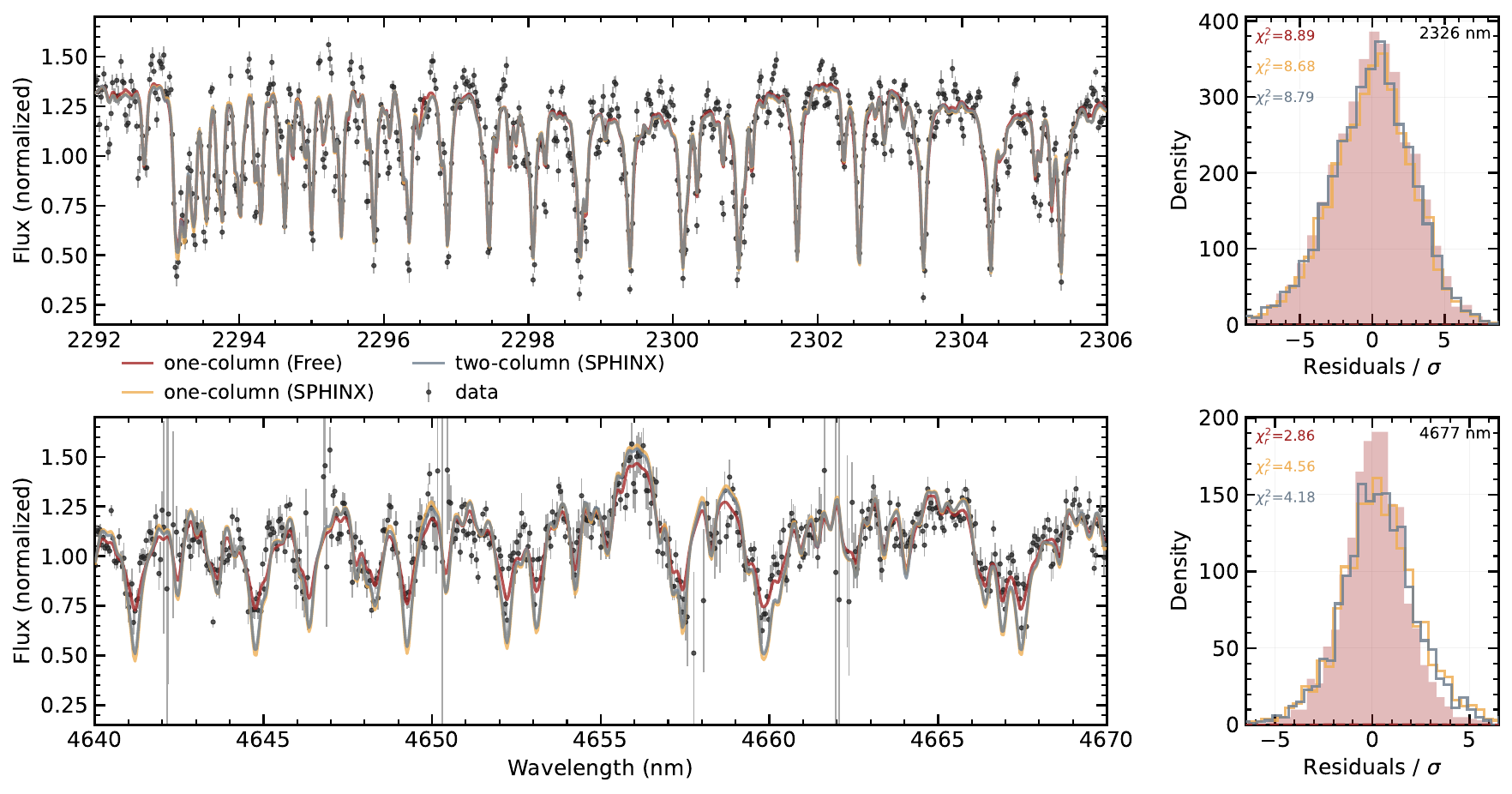}
    \caption{Comparison of the flexible one-column (red), one-column SPHINX (orange), and prior-driven two-column SPHINX (gray) retrievals. \textit{Left:} observed spectra (black) and model fits for representative \Kband{} (top) and \Mband{} (bottom) segments. \textit{Right:} residual distributions \((d_i-m_i)/\sigma_{\mathrm{eff},i}\) for a single order in each band, centered at $2.326$ and $4.677\,\micron$. Reduced $\chi^2$ values are annotated on the histograms. All three models fit the \Kband{} similarly, but the SPHINX models leave broader \Mband{} residuals, suggesting that the SPHINX models are not able to fit the \Mband{} data (jointly with the \Kband{} data) as well as the flexible one-column model.}
    \label{fig:residual_distributions}
\end{figure}

\begin{figure}
    \centering
    \includegraphics[width=\textwidth]{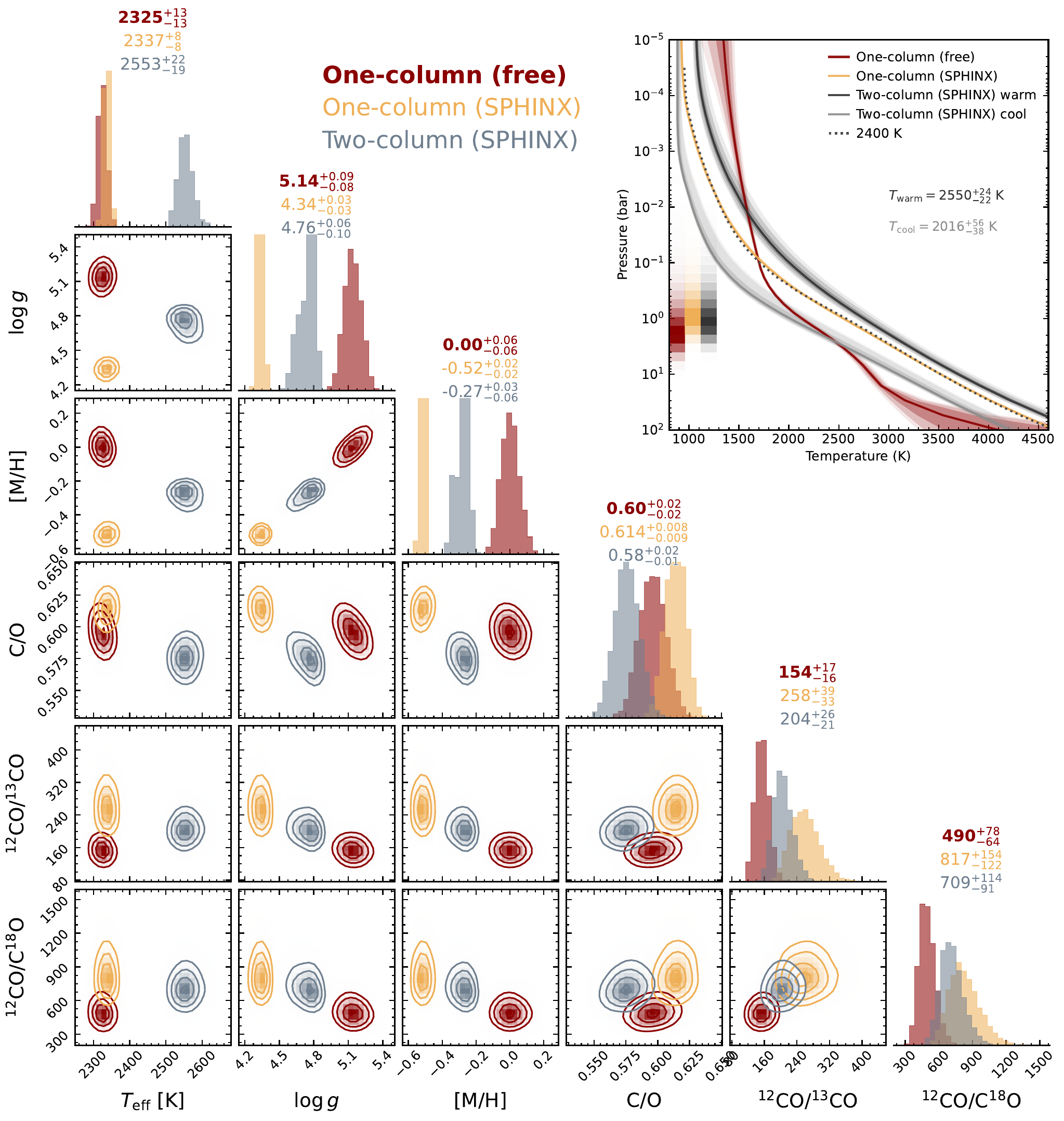}
    \caption{Posterior comparison of the flexible one-column (red), one-column SPHINX (orange), and prior-driven two-column SPHINX (blue-gray) retrievals. Diagonal panels show marginalized posteriors; off-diagonal panels show joint distributions. \textit{Inset:} retrieved pressure--temperature profiles, with warm and cool components shown separately for the two-column model. Median values and 16th--84th percentiles are annotated for each parameter. \(T_{\mathrm{eff}}\) is derived from bolometric flux integration.}
    \label{fig:two_column_cornerplot}
\end{figure}

This appendix expands on the exploratory two-column retrieval outlined in Sec.~\ref{sec:two_column}.
Initial tests with the flexible pressure--temperature parameterization of Sec.~\ref{sec:analysis} did not favor a two-column solution, because the single-column profile already has sufficient freedom to fit the data.
We therefore adopt SPHINX-II profiles to ask whether a linear combination of two physically motivated atmospheres can reproduce the flexible single-column fit (Figs.~\ref{fig:residual_distributions} and~\ref{fig:two_column_cornerplot}).
In preliminary runs without informative priors, a single-column solution with \(T_{\mathrm{eff}}\approx 2300\)~K was preferred and \(f_{\mathrm{cool}}\) was driven to the lower edge of its prior.
To explore a genuine two-column solution we impose priors that enforce a non-zero cool contribution:
\(T_{\mathrm{warm}} \sim \mathcal{N}(2550\,\mathrm{K},\,50\,\mathrm{K})\),
\(f_{\mathrm{cool}} \sim \mathcal{N}(0.40,\,0.10)\),
and
\(\Delta T_{\mathrm{eff}} \sim \mathcal{N}(400\,\mathrm{K},\,100\,\mathrm{K})\),
with \(T_{\mathrm{cool}} = T_{\mathrm{warm}} - \Delta T_{\mathrm{eff}}\).
These priors are motivated by low-resolution and photometric work, in particular the JWST analysis of \citet{rathckeStellarContaminationCorrection2025}, who prefer a cool component at \(\sim\)2000~K, a warm component at \(\sim\)2600~K, and a near-equal covering fraction.
As in that study, our best-fit cool component approaches the lower edge of the prior set by the coolest SPHINX-II model at 2000~K.
For comparison, we also run a one-column SPHINX retrieval with the same profile interpolation but a single component characterized by \(T_{\mathrm{eff,SPHINX}}\), alongside the flexible one-column retrieval of Sec.~\ref{sec:analysis} and the two-column SPHINX model.

The residual distributions are similar in the \Kband{} across all three models (Fig.~\ref{fig:residual_distributions}), whereas the largest differences appear in the \Mband{}.
There the two-column model provides a modest improvement over the one-column SPHINX fit but remains substantially worse than the flexible retrieval.
The statistical preference for the flexible model is therefore driven mainly by the \Mband{} fit quality, suggesting that the SPHINX PT profiles are not able to adequately describe the \Kband{} and \Mband{} data simultaneously.

Posterior comparisons (Fig.~\ref{fig:two_column_cornerplot}) show that the three retrievals span a wide range in \(T_{\mathrm{eff}}\), \(\log g\), and \([\mathrm{M/H}]\), where \(T_{\mathrm{eff}}\) is derived from bolometric flux integration as in Table~\ref{tab:trappist1_bestfit}.
The flexible and one-column SPHINX models yield similar effective temperatures, \(T_{\mathrm{eff}} = 2325^{+13}_{-13}\) and \(2337^{+8}_{-8}\)~K, whereas the two-column SPHINX retrieval gives \(T_{\mathrm{eff}} = 2553^{+22}_{-19}\)~K---closer to SED-based values despite comparable \Kband{} fit quality.
The SPHINX models also prefer substantially lower surface gravities (\(\log g = 4.34^{+0.03}_{-0.03}\) and \(4.76^{+0.06}_{-0.10}\)) and sub-solar metallicities (\([\mathrm{M/H}] = -0.52^{+0.02}_{-0.02}\) and \(-0.27^{+0.03}_{-0.06}\)) relative to the flexible retrieval (\(\log g = 5.14^{+0.09}_{-0.08}\), \([\mathrm{M/H}] = 0.00^{+0.06}_{-0.06}\)).
We attribute this degeneracy to a mismatch in the upper atmosphere: the SPHINX-II profiles are cooler than our flexible solution at low pressures, producing deeper CO absorption that is partially compensated by lower metallicity.
By contrast, C/O changes little with the temperature-structure parameterization (\(0.60^{+0.02}_{-0.02}\), \(0.614^{+0.008}_{-0.009}\), and \(0.58^{+0.02}_{-0.01}\) for the three models).
The isotopic ratios do not: the SPHINX models fit the \Mband{} poorly, where the isotopologue constraints are strongest (Fig.~\ref{fig:residual_distributions}), so we discard their discrepant \(\cisoratio\) and \(\oisoratio\) values rather than fold them into a systematic error budget.

\end{document}